\documentclass[fleqn,usenatbib]{mnras}

\usepackage{newtxtext,newtxmath}

\usepackage[T1]{fontenc}
\usepackage{soul}

\DeclareRobustCommand{\VAN}[3]{#2}
\let\VANthebibliography\thebibliography
\def\thebibliography{\DeclareRobustCommand{\VAN}[3]{##3}\VANthebibliography}

\usepackage{graphicx}	
\usepackage{amsmath}	
\usepackage{xspace}
\usepackage[dvipsnames]{xcolor}

\newcommand{\kmax}{ k_{\rm max} }
\newcommand{\hMpc}{ h^{-1}{\rm Mpc} }
\newcommand{\ihMpc}{ \,h\,{\rm Mpc}^{-1} }

\newcommand{\hMsun}{ h^{-1}{\rm M_{ \odot}}}
\newcommand{\om}{ \Omega_{\rm m} }
\newcommand{\dl}{ \delta_{\rm L}(\pmb{q}) }
\newcommand{\dg}{ \delta_{\rm g} }
\newcommand{\lamsat}{ \lambda_{\rm sat} }
\newcommand{\fsat}{ f_{\rm sat} }
\newcommand{\nseries}{{\tt Nseries}}

\newcommand{\qq}{ \pmb{q} }
\newcommand{\xx}{ \pmb{x} }
\newcommand{\sv}{ \pmb{s} }

\usepackage[dvipsnames]{xcolor}

\usepackage[normalem]{ulem}

\title[BACCO analysis of SDSS-III BOSS]{Cosmological constraints from the full-shape galaxy power spectrum in SDSS-III BOSS using the BACCO hybrid Lagrangian bias emulator}

\author[Pellejero Ib\'a\~nez et al.]{
Marcos Pellejero Ib\'a\~nez$^{1}$\thanks{E-mail: mpelleje@roe.ac.uk},
Raul E. Angulo$^{2,3}$\thanks{E-mail: reangulo@dipc.org},
and John A. Peacock$^{1}$
\\
$^{1}$Institute for Astronomy, University of Edinburgh, Royal Observatory, Blackford Hill, Edinburgh, EH9 3HJ, UK\\
$^{2}$Donostia International Physics Center (DIPC), Paseo Manuel de Lardizabal 4, 20018 Donostia-San Sebastian, Spain\\
$^{3}$IKERBASQUE, Basque Foundation for Science, E-48013, Bilbao, Spain
}

\date{Accepted XXX. Received YYY; in original form ZZZ}

\pubyear{\the\year{}}

\begin{document}
\label{firstpage}
\pagerange{\pageref{firstpage}--\pageref{lastpage}}
\maketitle

\begin{abstract}
We present a novel analysis of the redshift-space power spectrum of galaxies in the SDSS-III BOSS survey. Our methodology improves upon previous analyses by using a theoretical model based on cosmological simulations coupled with a perturbative description of the galaxy-matter connection and a phenomenological prescription of Fingers of God. This enables a very robust analysis down to mildly non-linear scales, $k\simeq 0.4\ihMpc$. We carried out a number of tests on mock data, different subsets of BOSS, and using model variations, all of which support the robustness of our analysis. Our results provide constraints on $\sigma_8$, $\Omega_m$, $h$, and $S_8 \equiv \sigma_8 \sqrt{\om/0.3}$. Specifically, we measure $\Omega_m=0.301\pm 0.011$, $\sigma_8=0.745^{+0.028}_{-0.035}$,  $h=0.705\pm 0.015$, and $ S_8 = 0.747^{+0.032}_{-0.039}$ when all the nuisance parameters of our model are left free. By adopting relationships among bias parameters measured in galaxy formation simulations, the value of $S_8$ remains consistent whereas uncertainties are reduced by $\sim20\%$. Our cosmological constraints are some of the strongest obtained with the BOSS power spectrum alone: they exhibit a $2.5-3.5\sigma$ tension with the results of the {\it Planck\/} satellite,
agreeing with the lower values of $S_8$ derived from gravitational lensing. However, the cosmological model preferred by {\it Planck\/} is still a good fit to the BOSS data, assuming small departures from physical bias priors and, therefore, cannot be excluded at high significance. We conclude that, at the present, the BOSS data alone does not show strong evidence for a tension between the predictions of $\Lambda$CDM for the high- and low-redshift Universe.

\end{abstract}




\section{Introduction}

The spatial distribution of galaxies encodes information about fundamental aspects of our Universe. For instance, the clustering amplitude as a function of scale depends on the abundance of different elements -- dark matter, baryons, and photons -- as well as the properties of the primeval Universe. Additionally, the clustering anisotropy induced by the peculiar velocities of galaxies, known as redshift-space distortions (RSD), depends on the growth rate of cosmic structures, which contains information about cosmological parameters and the law of gravity. 

This rich source of information has motivated large spectroscopic surveys, which map the 3-dimensional distribution of up to hundreds of millions of galaxies. It is not an exaggeration to say that a revolution in cosmology is on the horizon. Recent images released from ESA’s EUCLID satellite (\citealt{Euclid2024}) show the ability of the mission to deliver unprecedented maps of the cosmos. Additionally, after only 1 year of observations, the Dark Energy Spectroscopic Instrument (DESI, \citealt{DESI}) already built the largest ever galaxy redshift catalogue. Over the next decade, these surveys will be complemented by a myriad of other experiments such as 4MOST (\citealt{4MOST}), SPHEREX (\citealt{SPHEREX}), and J-PAS (\citealt{miniJPAS}).

An essential aspect of a successful exploitation of clustering data is accurate theoretical models. On the one hand, the spatial distribution of galaxies can be described with analytical approaches. These models are based on perturbative solutions of the relevant evolution equation for matter and for the connection between galaxies and the matter field (see the reviews by \citealt{Bernardeau_2002} and \citealt{Desjacques2018}). Although developments, such as the Effective Field Theory of Large-Scale Structure (EFT, e.g. \citealt{Baumann_2012,Carrasco_2014,Cabass_2023}), have considerably improved the accuracy and range of validity of such models, these approaches are only valid on large scales. On the other hand, in recent years it is becoming feasible to make direct use of cosmological simulations in performing analyses of redshift-space clustering data. These approaches have mostly been based on assuming Halo Occupation Distribution models, where the abundance of galaxies is primarily given by the mass of the host halo (see e.g. \citealt{SimBig_2022,knnpaper,Cuesta-Lazaro_2023}). Although these models can predict the clustering down to much smaller scales than e.g. EFT and therefore are potentially much more constraining, they make rigid assumptions about galaxy formation physics, which might not apply to the real universe. This could bias cosmological inferences, especially for samples with complex selection criteria.

An alternative family of theoretical models is given by hybrid methods (\citealt{Modi_2020}), which combine the accuracy of numerical simulations with the general formulation of pertubative models. Specifically, perturbative approaches are used to describe in full generality the large scale connection between galaxies and the linear matter field in Lagrangian coordinates. This Lagrangian ``biased'' field is then mapped to Eulerian space, and subsequently to redshift space (\citealt{PellejeroIbanez2022}), with $N$-body simulations. Therefore, we can take advantage of the accuracy of numerical simulations to increase the performance of perturbative models. Recently, this model has been used in studies of the 21-cm field in \cite{Baradaran_2024}.

In this paper, we will employ the ``BACCO hybrid Lagragian bias'' emulator that implements a 2nd-order Lagrangian bias expansion on top of the BACCO simulation suite \citep{Angulo_2021}. We have developed and validated our emulator in several previous papers. In \cite{ZennaroAnguloPellejero2021}, we showed that our approach describes the clustering of mock galaxies down to quasi-nonlinear scales. Then, in \cite{Zennaro_2022} we validated it against thousands of physical galaxy mocks built with various flavours of subhalo abundance matching. In parallel, in \cite{PellejeroIbanez2022} we extended and validated the hybrid-bias model for redshift-space clustering, and in \cite{PellejeroIbanez_2024} we explored its combination with field-level emulators. Finally, in \cite{PellejeroIbanez2023} we built the redshift-space model for thousands of different sets of cosmological parameters that included massive neutrinos and dynamical dark energy. With such data, we trained a neural-network emulator that predicts the monopole, quadrupole, and hexadecapole of the power spectrum for biased tracers as a function of cosmology. Our emulators have been further tested and compared with other approaches in preparation for the EUCLID and Rubin LSST surveys \citep{Pezzotta_2024,Nicola_2024}, and in the ``Beyond-2point'' challenge \citep{Beyond-2pt_2024}.

The goal of this paper is to present the first application of our redshift-space emulator to observational data. For this, we will use public measurements of the monopole, quadrupole, and hexadecapole of the redshift-space power spectrum of galaxies measured in SDSS-III BOSS \citep{BOSS2017}. This is the largest publicly-available redshift galaxy survey, with a volume of nearly 6 $[h^{-1}{\rm Gpc}]^3$ and over one million galaxies. 

We will show that our results provide cosmological constraints in broad agreement with previous analyses using the same BOSS datavector and either perturbation theory or simulation-based models. On the other hand, our analysis delivers 30\% tighter constraints on $\sigma_8$ with respect to pertubative approaches, which is a consequence of the unique features of our modelling. 

Although our constraints on $\Omega_m$ and $H_0$ are consistent with the recent analysis of the BAO feature in the 1st year of data in DESI \citep{DESIBAOCosmo}, our marginalised constraints on $S_8$ differ by $2.5-3.5\sigma$ with respect to those from {\it Planck\/}. Likewise, our $S_8$ constraint is in tension with the results from CMB lensing and cluster counts performed by the ACT \citep{ACTCosmo2024} and eROSITA \citep{eROSITA_2024} collaborations, respectively. However, our measurements are consistent with other low-redshift cosmological probes, such as joint analyses of lensing data and cluster abundances from the SPT survey \citep{SPTDESHST2024}, other full shape analysis \citep{ChenVlahWhite_2022,Philcox_2022,Ivanov_2023}, and joint analyses of lensing data and galaxy clustering \citep{Chen_2022,Chen_2024}. Despite this tension, we note that the cosmological parameter set preferred by {\it Planck\/} represents a very good description of the BOSS data, which suggests that the current BOSS data do not show significant enough evidence for a fundamental tension with the high-$z$ Universe within the standard $\Lambda$CDM model.

Our paper is organised as follows. We start in \S\ref{sec:data} by briefly describing the characteristics of the BOSS measurement, and, in \S\ref{sec:pipeline}, recapping our model. In \S\ref{sec:nseries} we then validate our inference pipeline against a suite of $N$-body mock catalogues resembling the BOSS data. Subsequently, in \S\ref{sec:results} we apply our pipeline to the redshift-space power spectrum of BOSS galaxies. We will first consider wide priors on the relevant cosmological, bias, and nuisance parameters. In a second stage, we will perform an exploratory analysis where we fix the relationship between the linear and high-order bias parameters according to the expectations from galaxy simulations. In \S\ref{sec:robustness}, we present several test of the robustness our results. Finally, we discuss the implications of our findings in \S\ref{sec:discussion}, and in \S\ref{sec:conclusions} we conclude.

\section{Galaxy Clustering Data}
\label{sec:data}

In this paper we will employ the redshift-space power spectrum of galaxies in  SDSS-III BOSS. Below we provide a summary of the  survey and the estimation of the power spectrum and the corresponding covariance matrix.

\subsection{BOSS survey} 
The dataset corresponds to the $12^{\rm th}$ data-release (DR12) of the Baryon Oscillation Spectroscopic Survey (BOSS, \citealt{BOSS-SDSSIII2013,BOSS-catalogs_2016}), a part of SDSS-III \citep{SDSSIII_2011}. BOSS encompasses 1,198,006 galaxies observed across two distinct regions of the sky: the Northern and Southern galactic caps (referred to as NGC and SGC, respectively). There are two galaxy selection criteria in BOSS: CMASS and LOWZ, which roughly correspond to high-redshift galaxies above a stellar mass threshold and low redshift galaxies above a magnitude limit. 

For the purposes of this analyses, we consider a division of each galactic cap only in terms of redshift: $z_1 \in [0.2, 0.5]$ and $z_3 \in [0.5, 0.75]$, with a median redshift of $0.38$ and $0.61$, respectively. Although each of these redshift ranges will contain a mixture of CMASS and LOWZ galaxies, we expect this not to affect our modelling given the generality of the bias expansion. The raw volume of each of these four subsamples is 1.7\,Gpc$^3$ and 3.3\,Gpc$^3$ for SGC, and 4.7\,Gpc$^3$ and 9\,Gpc$^3$ for the NGC \citep{BOSS2017}, to give a total of 18.7\,Gpc$^3$ (approximately 6\,$[h^{-1}{\rm Gpc}]^3$). Therefore, we expect the high-redshift NGC to be the sample with the largest statistical power. Conversely, the low-redshift SGC is expected to display the weakest constraining power.


\subsection{Power spectrum estimation}

We employ the power spectrum of galaxies in each of the 4 samples, as measured by \cite{Philcox_2022}. The statistic is computed using an ``unwindowed'' estimator, which aims  to deliver the power spectrum without the effect of the survey mask, easing the comparison with theoretical models. In practice, the estimators are obtained by defining the large-scale likelihood for the galaxy survey and subsequently optimising analytically for the desired statistics, analogous to methodologies employed in early analyses of CMB and Large-Scale Structure (LSS). 

The specific formulation of the estimator used by \cite{Philcox_2022} was proposed in \cite{Philcox_2021,Philcox_2021b} and \cite{Philcox_2024}. These authors approximate the pixel covariance using a diagonal (FKP-like) structure with $P_{\rm fkp} \approx 10^4\,[h^{-1}{\rm Mpc}]^3$, while still fully considering the survey's geometry. In addition, the measurements are corrected by the impact of observational systematic errors such as galaxy-star separation, redshift failures, spectroscopic fibre collisions, and observing conditions. The public measurements\footnote{Publicly available at \url{https://github.com/oliverphilcox/Spectra-Without-Windows}} provide the monopole, quadrople, and hexadecapole of the power spectrum with a resolution of $\Delta k = 0.005\ihMpc$ up to $k=0.41\ihMpc$ which is given by the Nyquist frequency of the Fourier grid employed $k_{\rm Nyq} = 0.45\ihMpc$. 

Note that the smallest scale provided by the public measurements will determine the wavenumbers we employ in our analysis, despite our theoretical model being, in principle, able to use even smaller scales ($\kmax \sim 0.6\,\ihMpc$), as shown in \cite{PellejeroIbanez2023}.

\subsection{Covariance Matrix}
\label{sec:covariance}

The covariance matrix of the power spectrum measurements was estimated using a set of 2048 \texttt{MultiDark-Patchy} mock catalogues \citep{Kitaura_2016,Rodriguez-Torres_2016}, publicly provided by the BOSS collaboration\footnote{Available at \url{https://data.sdss.org/sas/dr12/boss/lss/}}. In particular, we use the window-free estimation by \cite{Philcox_2022}. 

Each \texttt{Patchy} catalogue is built by creating a nonlinear density and velocity field with a fast but approximate gravity solver, which is then calibrated to reproduce the results from a high-resolution $N$-body simulation. The \texttt{Patchy} catalogues have a flexible model for the galaxy-matter connection, whose free parameters were set to reproduce the observed redshift-space power spectrum in BOSS. Finally, mock galaxies are selected to display the same observational setup as BOSS in terms of its window mask and weights due fibre collisions.

We note that the \texttt{Patchy} mocks have only been extensively validated for $k<0.3\ihMpc$, yet, we will employ them until $k \simeq 0.4\hMpc$. We justify this by noting that, for the number density and volume of BOSS, the diagonal elements of the covariance matrix are dominated by the Gaussian terms, whereas off-diagonal elements are dominated by mode-coupling caused by the survey mask \citep{Lippich18, Colavincenzo18, Blot19}. Therefore, since all these elements have been carefully incorporated into the \texttt{MultiDark-Patchy} mocks, we expect our covariance matrix to provide a reasonably accurate description of the uncertainties over the full range of scales we will employ. In the future, we will study the dependency of our results with alternative methods to estimate covariance matrices \citep{Balaguera2019b, Balaguera2019a,Pellejero2020a, Sinigaglia2020, Kitaura2020,Ereza2023}.

Although \texttt{Patchy} represents an adequate method to estimate covariance matrices, its inherent approximations might compromise the accuracy of the mock multipoles, specially at $k > 0.2 \ihMpc$. For this reason, we will validate our inference pipeline using a different suite of mock catalogues (see \S\ref{sec:nseries}). 

\section{Theoretical modelling}
\label{sec:pipeline}

In this section, we provide details about our cosmology inference pipeline. In \S\ref{sec:model}, we discuss our modelling for the redshift-space clustering of biased tracers, including the hybrid bias expansion and how we emulate our predictions. In \S\ref{sec:inference}, we provide details on our Bayesian inference, including the definition of the parameter space, the form of the likelihood, and the sampler. 

\subsection{The galaxy power spectrum in redshift space}
\label{sec:model}
\subsubsection{Hybrid Lagrangian bias expansion}
\label{sec:bias_expansion}

The modelling of clustering statistics within the "hybrid" approaches, as outlined in \cite{Modi_2020} and extended in \cite{PellejeroIbanez2022}, involves two key components: i) a mapping from Lagrangian ($\qq$) to Eulerian ($\xx$) space given by the displacement field, $\pmb{\psi}(\qq)$, measured in $N$-body simulations, and ii) a functional relationship in Lagrangian space between the linear overdensity field, $\dl$, and the galaxy density field, $\dg(\qq)$. 

The relationship between matter and galaxies, $F[\dl]$, is typically referred to as the bias model. In our approach, we adopt a 2nd-order perturbative expansion for $F(\qq)$, as described in \cite{Matsubara2008} and \cite{Desjacques2018}, which weighs the significance of various Lagrangian fields in depicting the density of tracers at a given $\qq$:

\begin{equation}
\begin{split}
     \delta_g(\qq) \equiv F[\delta_{\rm L}(\qq)] = & 1 + b_1\delta_{\rm{L}}(\qq) + b_2 \left( \delta^2_{\rm{L}}(\qq)-\langle\delta^2_{\rm{L}}(\qq)\rangle \right) \\ & + b_s \left( s^2(\qq) - \langle s^2(\qq)\rangle \right) + b_{\nabla^2} \nabla^2 \delta_{\rm{L}}(\qq) \; .
	\label{eq:model}
\end{split}
\end{equation}

\noindent Here, brackets $\langle\cdot\rangle$ denote volume averages, and $s^2 \equiv s_{ij}s^{ij}$ where $s_{ij}$ is the traceless tidal tensor. Thus, $s^2 = \partial_i\partial_j\phi(\qq) - 1/3\delta^{\rm{K}}_{ij}\delta_{\rm{L}}(\qq)^2$, with $\phi(\qq)$ representing the linear gravitational potential. 

Therefore, the overdensity field of galaxies in Eulerian space,  $\dg(\xx)$, reads:

\begin{equation}
\begin{split}
    1+\dg(\xx) = \int {\rm d}^3q \, F[\dl] \, \delta_{\rm D}(\xx-\qq-\pmb{\psi}(\qq)),
    \label{eq:galmapp}
\end{split}
\end{equation}

\noindent where $\delta_{\rm D}$ is a Dirac's delta. We highlight that our implementation of the hybrid bias expansion has undergone extensive testing. For instance, in \cite{Zennaro_2022}, we built thousands of mock catalogues based on the {\it SubHalo Abundance Matching extended} model \citep[SHAMe][]{ContrerasAnguloZennaro2020AB,ContrerasAnguloZennaro2020} with different parameter sets, selection criteria, number densities, redshifts, and cosmologies. We then demonstrated that our approach was able to accurately fit the galaxy power spectra down to $\kmax \simeq 0.7\ihMpc$. Given the success of the model in real space, recently we extended the hybrid approach to model the intrinsic alignment of galaxy shapes \citep{Maion_2023}.

To compute the galaxy overdensity in redshift space, $\dg^z$, we include an additional term in the displacement field to account for the effect of peculiar velocities \citep{PellejeroIbanez2022}:

\begin{equation}
\begin{split}
    \pmb{\psi}(\qq) \rightarrow \pmb{\psi}(\qq) + \frac{\hat{\xx}_z(\qq) \cdot \pmb{v}[\xx(\qq)]}{aH} \, \hat{\xx}_z(\qq),
    \label{eq:galDispRSD}
\end{split}
\end{equation}

\noindent where $\hat{\xx}_z({\qq})$ represents the unit vector along the line-of-sight, $a$ denotes the scale factor, and $H$ is the Hubble parameter. The velocity field $\pmb{v}$ is built directly from the peculiar velocities measured for haloes and matter in $N$-body simulations. Thus, our model inherits in full the nonlinearity of velocities and of the mapping between real-, $\xx$, and redshift-space coordinates, $\sv$.

We note that we cannot simply use $v(\xx)$ as the velocity of matter in simulations because we expect galaxies to sample a coarsed-grained version of the velocity field. For instance, central galaxies will be mostly at rest relative to their host halo, whereas satellite galaxies could display larger or smaller velocity dispersion than typical particles in a halo, owing to the details of galaxy formation physics \citep{Orsi_2018,Alam_2021}. Therefore, we define $\pmb{v}(\xx) = \pmb{v}_{\rm matter}(\xx)$ if $\xx$ is outside a halo and $\pmb{v}(\xx) = \pmb{v}_{\rm{halo}}(\xx)$ if $\xx$ is found within a halo. To model the role of intra-halo satellite velocities, we include two additional free parameters that describe the fraction of satellite galaxies, $f_{\rm sat}$, and their typical velocity dispersion $\lamsat$:

\begin{equation}
    \dg^s(\sv) \rightarrow \dg^s(\sv) \boldsymbol{\ast}_z \left[ (1-\fsat)\, \delta_{\rm{D}}(s_z) +\fsat \,\exp \left( {-\lamsat s_z} \right) \right]. 
\end{equation}

\noindent where $\boldsymbol{\ast}_z$ denotes a convolution along the line-of-sight direction. In \cite{PellejeroIbanez2022} we extensively tested this model for different SHAMe catalogues in redshift space (including one that mimics the CMASS sample), finding it describes the clustering of galaxies down to $\kmax \sim 0.6\ihMpc$ at the precision of the BOSS survey. We also tested this model in a {\it blind} manner within the ``Beyond 2-pt Challenge'' \citep{Beyond-2pt_2024}, where our approach delivered unbiased and among the strongest constraints on $\sigma_8$. 

Although the model provides predictions at the field level  \citep{PellejeroIbanez_2024}, in this paper we are only interested in the redshift-space power spectrum which is typically characterised by a multipole expansion:

\begin{equation}
P_{\ell}(k) = \frac{2\ell+1}{2}\int_{-1}^{1}d\mu\,\mathcal{P}_\ell(\mu)\,\langle |\dg^s(k,\mu)|^2) \rangle,
\label{eq:pkpoles}
\end{equation}

\noindent where $k\equiv |\pmb{k}|$, $\mu \equiv \pmb{k} \cdot \pmb{k_z}$, and $\mathcal{P}_\ell(\mu)$ are Legendre  polynomials.

Finally, we include two stochastic terms in the power spectrum monopole, $\ell=0$, to account for discreteness noise and small-scale physics not included in our bias model: $\epsilon(k) = 1 /\bar{n} \, (\epsilon_1+\epsilon_2k^2)$, where $\bar{n}$ is the mean number density of galaxies in the sample. Note that in \cite{PellejeroIbanez2022} we tested the need to include $\mu$-dependent stochastic terms, finding them not required for the range of scales and level of accuracy of BOSS. This does not imply the term is absent, rather, it indicates that the velocity from halos in the simulation, combined with the FoG free parameters, accounts for this dependency.

\subsubsection{Emulation}
\label{sec:emulation}

For our analysis, we will employ the redshift-space hybrid bias emulator presented in \cite{PellejeroIbanez2023}. Below, we will recap its main features.

For our model to make predictions as a function of cosmological parameters, we require displacement and velocity fields measured in $N$-body simulations for multiple cosmologies. Obtaining such predictions is a computationally-challenging task, thus, we resort to the BACCO suite of high-resolution simulations (initially introduced in \citealt{Angulo_2021}), which can be transformed to span thousands of different cosmologies by employ a cosmology-rescaling algorithm \citep{AnguloWhite2010,Zennaro_2019,Contreras_2020}. 

We highlight that, in addition to the large volume of BACCO simulations ($V\simeq 8 {\rm Gpc}^3$ with $4320^3$ particles of $\sim 10^{9}\hMsun$ mass), these simulations have ``fixed-and-paired'' initial conditions, which significantly reduced the cosmic variance errors in the simulated displacement and velocity fields \citep{Angulo2016,Maion2022}.

We rescaled the BACCO suite to 4000 combinations of cosmological parameters and redshifts, over a hyperspace in the cosmological parameters $\{\Omega_{\rm m} , \sigma_8 ,  \Omega_b, n_s, h, M_{\nu} \,[{\rm eV}], w_{0},  w_{a}\}$ that roughly spans 10 times the uncertainty delivered by the analysis of {\it Planck\/} data \citep{PlanckCosmo2020}. Although our emulator provides predictions as a function of neutrino mass, $M_{\nu}$, and dynamical dark energy, $w(a) \equiv w_0 + (a - 1)\,w_a$, we will perform our current analysis within the minimal $\Lambda$CDM model: $M_{\nu} = 0$eV, $w_0=-1$, and $w_a=0$\footnote{Note we could have fixed $M_{\nu} = 0.06$eV, but this had a minimal impact in our results.}.

At each parameter set, we build the hybrid model predictions by advecting each of the 5 terms in Eq.~\ref{eq:model} to redshift space, and compute all the 15 cross-power spectra that enter in Eq.~\ref{eq:pkpoles}. To reduce statistical noise as much as possible, we repeated and average our calculations over the 3 line-of-sight directions, $\{\hat{x}, \hat{y}, \hat{z}\}$.  

These measurements provided a library of power spectrum multipoles at different cosmologies that allows for emulation via a neural network. \cite{PellejeroIbanez2023} demonstrated that the typical accuracy for the monopole, quadrupole, and hexadecapole is 0.5\%, 1\%, and 10\%, respectively. These are subdominant compared to the statistical uncertainties in BOSS, where the smallest uncertainties for the NGC subsample are $3-4\%$ for the monopole, $20\%$ for the quadrupole, and $40\%$ for the hexadecapole.

\subsubsection{The Alcock-Paczynski effect}

The clustering of galaxies would appear anisotropic if the cosmology assumed to transform redshifts and angles into comoving separations is incorrect. 

To account for this effect, known as Alcock-Paczynski distortion \citep{AlcockPaczynski_1979}, we first construct a 2D power spectrum, $P(k,\mu)$, from the three multipoles that we emulate. Then, we transform our measurements to how they would have appeared if we had assumed the same cosmology as the BOSS collaboration \citep{Ballinger_1996}:

\begin{eqnarray}
\label{eq:AP}
    P_{\rm fid}(k,\mu) &=& \left(\frac{D_{A,{\rm fid}}}{D_{A}}\right)^2\left (\frac{H}{H_{\rm fid}}\right) P(k',\mu'),\\
 k^{\prime}  &\equiv&  k\,\left[ \left(\frac{H}{H_{\rm fid}}\right)^2\mu + \left(\frac{D_{A,{\rm fid}}}{D_{A}}\right)^2(1-\mu^2) \right]^{\frac{1}{2}} \nonumber\\
 \mu^{\prime}  &\equiv&  \mu \left(\frac{H}{H_{\rm fid}}\right) \left[ \left(\frac{H}{H_{\rm fid}}\right)^2\mu + \left(\frac{D_{A,{\rm fid}}}{D_{A}}\right)^2(1-\mu^2) \right]^{-\frac{1}{2}} \nonumber
\end{eqnarray}

\noindent where $H$ and $D_A$ are the Hubble parameter and the angular diameter distance at the median redshift of each sample. The "fid" subscript refers to quantities computed adopting the fiducial cosmology in the BOSS survey.

At each likeihood evaluation, we compute the transformation at the corresponding cosmological parameters. We then compute $P_{\rm fid}$ and the respective power spectrum multipoles using Eq.~\ref{eq:pkpoles}. 

\subsection{Parameter Inference}
\label{sec:inference}

\subsubsection{Likelihood and Sampling}
\label{sec:likelihood}

Due to the central limit theorem, we will assume that the likelihood of observing a set of power spectrum multipoles, $\vec{P}_{\ell}=\{P_0,P_2,P_4\}$, has a multivariate Gaussian form:

\begin{eqnarray}
\label{equation:clust_like}
 \log \mathcal{L} = -\frac{1}{2} \left[\vec{P}_\ell(k) - \vec{P}_\ell^{\rm bacco}(k)\right]^{\rm{T}}\,C_{\rm{BOSS}}^{-1}\,\left[\vec{P}_\ell(k) - \vec{P}_\ell^{\rm bacco}(k)\right] \nonumber \\
 - \frac{1}{2}\log|C_{\rm{BOSS}}|- \frac{N}{2}\log(2\pi)\, , 
\end{eqnarray}

\noindent where $\vec{P}_\ell^{\rm bacco}$ are the multipoles evaluated with the model described in \S\ref{sec:bias_expansion}, and $C_{\rm{BOSS}}$ is the covariance matrix  estimated using \texttt{Patchy} mocks \citep{Kitaura_2016}, as detailed in \S\ref{sec:covariance}. When combining samples, we assume them to be independent from each other, i.e. the full covariance matrix is block diagonal. 

We compute the likelihood and credibility intervals on our parameters using a nested sampler algorithm, as implemented in the publicly available code \texttt{MULTINEST}\footnote{\url{https://github.com/farhanferoz/MultiNest}}  \citep{multinest1,multinest2,multinest3}. Our analysis uses an evidence tolerance of 0.1 and a number of live points of 250. We tested with more stringent setups (live points of 400) for our fiducial result and found no differences, as opposed to \cite{Lemos_2023}. This is likely due to the reduced parameter space in comparison to their work.

After evaluating the 15 emulators, including the FoG and noise effects and performing the AP transformations, one likelihood evaluation takes approximately 1 second. This is a reasonable time for running chains with the nested sampler, provided we restrict ourselves to parameter spaces smaller than about 30 parameters. For larger parameter spaces, we would need to speed up the likelihood computation, for example, by including the AP parameters as part of the emulated parameters. We will explore these options in future work.

\subsubsection{Parameter space}
\label{sec:parameters}

\begin{table}
  \centering
  \begin{tabular}{ccc} 
 \hline
 \hline
 \multicolumn{3}{c}{\bf Cosmology} \\
 \hline
 & $\Omega_{\rm m}\,h^2$  & $\mathcal{U}(0.08,0.256)$ \\
 & $\sigma_8$  & $\mathcal{U}(0.65, 0.9)$ \\
& $h$             & $\mathcal{U}(0.6, 0.8)$ \\
& $\Omega_b\,h^2$  &$\mathcal{N}(0.02236, 0.00014)$\\
& $n_s$  & $\mathcal{N}(0.9649, 0.0038)$\\
\hline\hline
 \multicolumn{3}{c}{\bf Galaxy-Matter connection} \\
 \hline
& $b_1$ & $\mathcal{U}(0,2)$ \\
& $b_2$ & $\mathcal{U}(-2,2)$  \\
& $b_{s^2}$ & $\mathcal{U}(-5,5)$ \\
& $b_{\nabla^2}/[{\rm Mpc}/h]^2$ & $\mathcal{U}(-5,5)$ \\

\hline \hline
\multicolumn{3}{c}{\bf Stochasticity} \\
\hline
& $\epsilon_1$  & $\mathcal{U}(-3,3)$ \\
& $\epsilon_2$  & $\mathcal{U}(-10,10)$ \\
\hline \hline
\multicolumn{3}{c}{\bf Redshift-Space Distortions}\\
\hline
& $\lambda_{\rm FoG}$ & $\mathcal{U}(0,1)$ \\
& $f_{\rm sat}$ & $\mathcal{U}(0,1)$ \\
    \hline
\end{tabular}

 \caption{\label{tab:priors} Free parameters and priors of the BACCO hybrid model. ${\mathcal U} (a,b)$ indicates a uniform distribution over the range $[a,b]$, and $\mathcal{N}(\mu, \sigma)$ a Gaussian distribution with mean $\mu$ and standard deviation $\sigma$. }
\end{table}

Our theoretical model, $P_\ell^{\rm bacco}$, contains 13 free parameters: 5 cosmological, 6 describing the galaxy-matter connection and its stochasticity, and 2 specifying the intra-halo velocity dispersion.
These parameters are listed in Table \ref{tab:priors}.

For the three main cosmological parameters we expect to be most constrained by our data $\{\Omega_m h^2, \sigma_8,  h\}$ we will adopt flat priors that coincide with the ranges provided by the emulator. For the two remaining cosmological parameters, $\{n_s$, $\Omega_b\}$ we will adopt informative Gaussian priors given by the marginalised posterior of CMB analyses. For all the remaining (nuisance) parameters, we will adopt flat and uninformative priors. 

We note that our bias parameters describe the average relation between galaxies in a given sample and the underlying Lagrangian density field. Therefore, in our fiducial analysis we have an independent set for each BOSS subsample. For instance, when combining the low-$z$ NGC and SGC measurements, we will have 21 free parameters: 5 cosmological and (2$\times$8=)16 independent bias parameters. 

It is interesting to note that for a given population, we do not expect the bias parameters to be completely independent. For instance, there are well established relations among bias parameters for dark matter haloes \citep[see][and references therein]{StueckerPB_2024}. Although these are not directly applicable to galaxies (due to assembly bias and because galaxy formation physics and selection criteria modulate the haloes entering the sample), the current generation of galaxy formation simulations limits the magnitude of the departures from the halo bias relations. Additionally, as our understanding of galaxy formation improves, these bias relations will become more robust and provide a clear path to incorporate galaxy formation physics to improve cosmological constraints. 

To explore this idea, we will complement our fiducial results with an analysis in which we express higher-order bias parameters as a function of the linear bias $b_1$. Explicitly, we will adopt the mean relationship for Lagrangian bias parameters reported by \cite{Zennaro_2022}:  

\begin{eqnarray}
\label{eq:param_range}
b_2(b_1)&=&0.01677 b_1^3-0.005116 b_1^2+0.4279 b_1 - 0.1635 \nonumber\\
b_{s^2}(b_1)&=& -0.3605 b_1^3+0.5649 b_1^2-0.1412 b_1 - 0.01318 \nonumber\\
\frac{b_{\nabla^2}(b_1)}{[{\rm Mpc}/h]^2}&=& 0.2298 b_1^3-2.096 b_1^2+0.7816 b_1 - 0.1545 \nonumber
\end{eqnarray}

\noindent which represent the mean over thousands of catalogues of different redshifts, selection criteria, underlying cosmology, and sample number densities. Importantly for consistency, these measurements were obtained by fitting the real-space galaxy power spectrum and the matter-galaxy cross correlation using the same hybrid-bias model of this paper. In \cite{PellejeroIbanez2022} we checked that these values were consistent between real- and redshift- space.

We will refer to this alternative analysis as ``Physical Bias Priors''. By adopting information about the relationship among bias parameters, we will  reduce the degrees of freedom in our model. This results in a more deterministic prediction for a given cosmological model, lower prior-volume effects, and thus stronger cosmological constraints (see e.g. \citealt{Ivanov_2024}). 

\section{Validation on mock catalogues}
\label{sec:nseries}

\begin{figure}
   \includegraphics[width=\columnwidth]{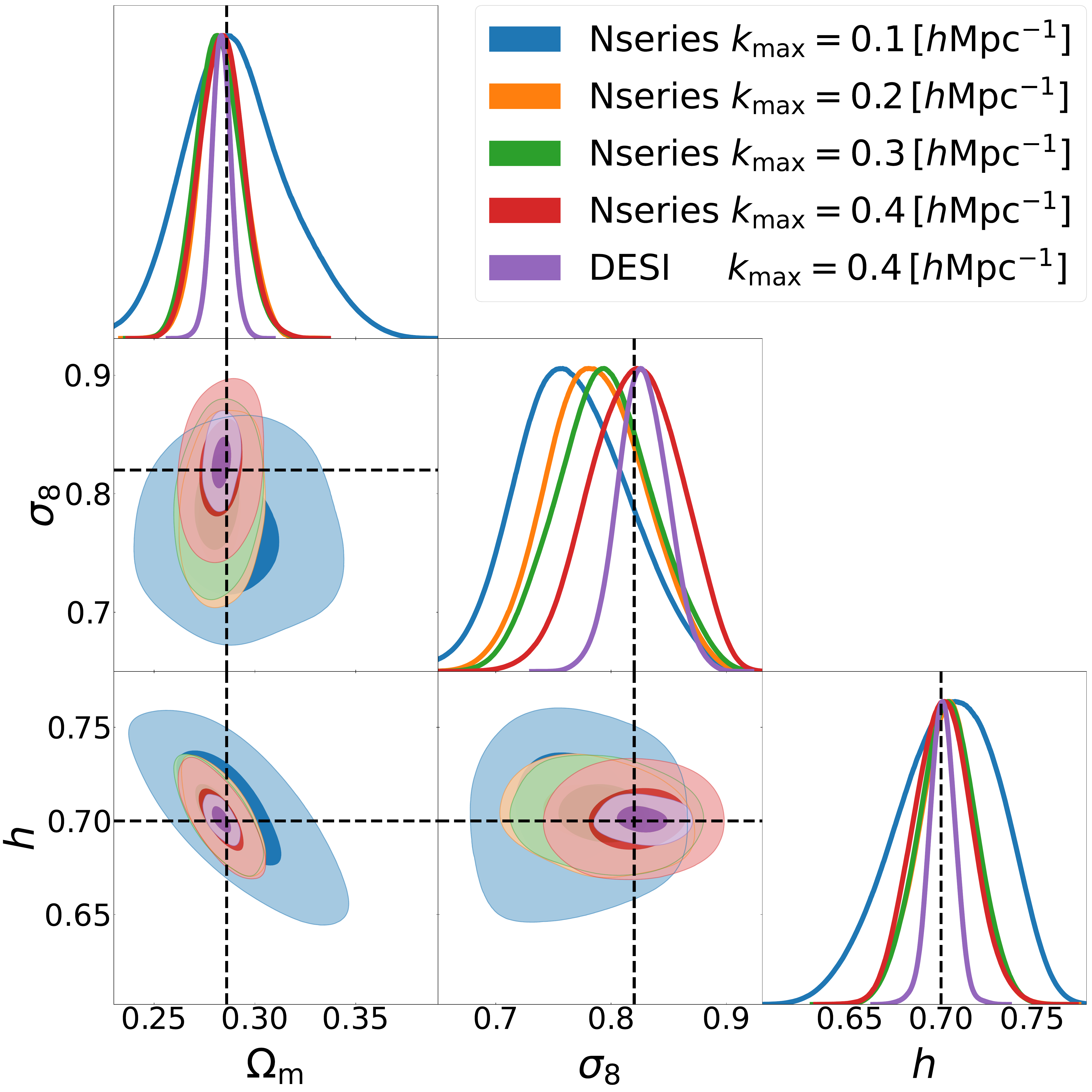}
   \caption{Validation of our cosmological inference pipeline in the \nseries\ suite of mock catalogues. We display the 68\% and 92\% confidence intervals on $\om$, $\sigma_8$, and $h$ obtained from applying our methodology to the monopole, quadrupole, and hexadecapole measured in the mocks, for three different values of $\kmax$ the maximum wavenumber included, as indicated by the legend. The synthetic datavector corresponds to the average of 84 pseudo-independent mocks built with an ``Halo Occupation Distribution´´ model that resembles galaxies in BOSS, and with a covariance matrix corresponding to the full BOSS volume. Additionally, we include as purple contours the results of assuming a covariance matrix corresponding to a 5-times larger volume, similar to that expected for the whole DESI sample ($V_{\rm eff} = 30 h^{-3}{\rm Gpc}^3$). Note that the regions displayed is smaller than that covered by our priors and were chosen to improve the clarity of the plot.
    \label{fig:Nseries}}%
\end{figure}

Before presenting the main results of this study, we validate our methodology by applying it to the \nseries mock catalogues\footnote{Available at \url{https://www.ub.edu/bispectrum/page11.html}}\citep{BOSS2017}.

\subsection{The \nseries\ simulation suite}

\nseries\ is a suite of 84 quasi-independent mock catalogues of the CMASS NGC sample of the BOSS survey. These mocks were generated from seven independent $N$-body simulations with $2500\,\hMpc$ a side and adopting the following cosmological parameters: $\om=0.286$, $\sigma_8=0.82$, $h= 0.7$, $n_s=0.97$. Mock galaxies were creating by using an HOD formalism, with parameters expected to describe the CMASS sample. Each simulation was projected into 12 different orientations adopting the same radial and angular selection function and fibre incompleteness as the BOSS data \citep{BOSS2017}. 


Although these mocks only encompass the CMASS NGC sample, spanning $0.43 < z < 0.7$ (with an effective redshift of $z_{\rm eff} = 0.56$), the substantial volume of the mocks and the detailed modelling of redshift-space distortions, allow us to test the ability of our model to deliver unbiased cosmological constraints. Given the differences with respect to the complete BOSS survey, a specific covariance matrix is required for the {\nseries}. This was measured with 2048 \texttt{Patchy} simulations adopting the window function of the \nseries\ mocks \citep{Philcox_2022}. Finally, the mocks were transformed assuming the fiducial cosmology: ${\Omega_{\rm m,fid}= 0.31,h_{\rm fid}= 0.676}$, which will test our treatment of geometric redshift-space distortions.


\begin{table*}
  \centering
  \begin{tabular}{cccccccccc} 
     \hline
     Survey &  $\om$ & $\sigma_8$ & $h$ & $\om h^2$ & $S_8$\\
     \hline
    BOSS-NGC & $0.301\pm 0.011$ & $0.745^{+0.028}_{-0.035}$ & $0.705\pm 0.015$ & $0.1498\pm 0.0046$ & $0.747^{+0.032}_{-0.039}$ \\
             & $(0.300)$ & $(0.724)$ & $(0.699)$ & $(0.146)$ & $(0.724)$ \\
    BOSS-SGC  & $0.279\pm 0.016$ & $0.748^{+0.045}_{-0.053}$ & $0.682^{+0.021}_{-0.025}$ & $0.1298\pm 0.0062$ & $0.722^{+0.048}_{-0.062}$ \\
             & $(0.295)$ & $(0.786)$ & $(0.684)$ & $(0.138)$ & $(0.780)$ \\
    BOSS + Physical Bias Priors  & $0.2983\pm 0.0086$ & $0.736\pm 0.025$ & $0.693\pm 0.013$ & $0.1431\pm 0.0028$ & $0.734\pm 0.028$ \\
             & $(0.300)$ & $(0.736)$ & $(0.694)$ & $(0.145)$ & $(0.737)$ \\
    \hline
    {\it Planck\/} & $0.315\pm0.0085$ & $0.812\pm0.0075$ & $0.6730\pm0.0061$ & $0.1425\pm0.0013$ & $0.832\pm0.016$ \\
     \hline
     \end{tabular}
    \caption{Mean and 68\% confidence intervals of the marginalised posterior distribution on cosmological parameters inferred from the BOSS data. The bottom row shows the cosmology inferred by the {\it Planck\/} collaboration. The results of BOSS assume {\it Planck\/} Gaussian priors on $\Omega_bh^2$ and $n_s$. The parameters not stated here correspond to a $\Lambda$CDM cosmology with massless neutrinos ($M_{\nu}=0$ eV) and optical depth at recombination $\tau=0.0952$. In parenthesis, we report on the values that maximise the likelihood of our the BOSS data given our model.}
  \label{tab:parameters_table}
\end{table*}

\subsection{Validation}

We validate our model and likelihood by inferring cosmological parameters from the power spectrum multipoles measured in the \nseries\ mocks. 

We use the mean over the 84 mocks as our synthetic data vector. We consider two cases for the uncertainties. First, we employ a covariance matrix corresponding to the total effective volume of BOSS ($V_{\rm eff} = 6 h^{-3}{\rm Gpc}^3$). Second, we consider a covariance matrix rescaled by a factor of 5, roughly corresponding to the full volume of a DESI-like survey,  ($V_{\rm eff} = 30 h^{-3}{\rm Gpc}^3$). We dub these cases as  ``Nseries'' and ``DESI'', respectively. The first case will validate the use of our pipeline to BOSS data, whereas the second analysis will provide a much more stringent test of our theoretical modelling.

We display our results in Fig.~\ref{fig:Nseries}. Our first finding is that we obtain unbiased constraints on the three main cosmological parameters at $\kmax=0.4\ihMpc$. In addition, the posterior distributions are consistent when only considering larger scales: the cosmological constraints are statistically compatible for all the values of $\kmax$ we consider. Additionally, we observe a progressive increase in  constraining power when including small scales, specially in $\sigma_8$. For instance, the $1\sigma$ region on $\sigma_8$ decreases by $\sim25\%$ when comparing $\kmax=0.1$ and $0.4\ihMpc$. On the other hand, uncertainties on $\Omega_{\rm m}$ and $h$ are fairly stable for $\kmax > 0.2\,\ihMpc$. As discussed in \cite{Pellejero2020a}, we do not expect very large gains beyond $\kmax = 0.2\ihMpc$ for the volume and number density of tracers in BOSS. Indeed, by using our methodology down to $0.4\ihMpc$, the Figure-of-Merit can increase by a factor of $1.1$ compared to the scales usually employed in similar analyses, $\kmax=0.2\ihMpc$. This is equivalent to having access to a survey volume $20\%$ times larger.

When considering the ``DESI'' case (purple contours), we see that our constraints on all parameters agree within the true cosmology (indicated by dashed lines) at the $1\sigma$ level. We note that these contours coincide with those from the Nseries analysis, which suggests that projection effects in the parameter space are not significant for $\kmax=0.4\ihMpc$. However, there can be projection effects at $\kmax=0.2\ihMpc$, as indicated by the small shifts in $\sigma_8$. Although not shown here, we tested with a DESI volume covariance, and the shifts reduced their statistical significance. We explore this further in \S\ref{sec:prior_volume}.


Therefore, we conclude that theoretical uncertainties in our model are negligible compared to statistical uncertainties in BOSS, and thus our pipeline should deliver robust and accurate parameter constraints using mildly non-linear scales up to $\kmax=0.4\,\ihMpc$.

\section{Results}
\label{sec:results}

   \begin{figure*}
   \centering
   \includegraphics[width=0.7\textwidth]{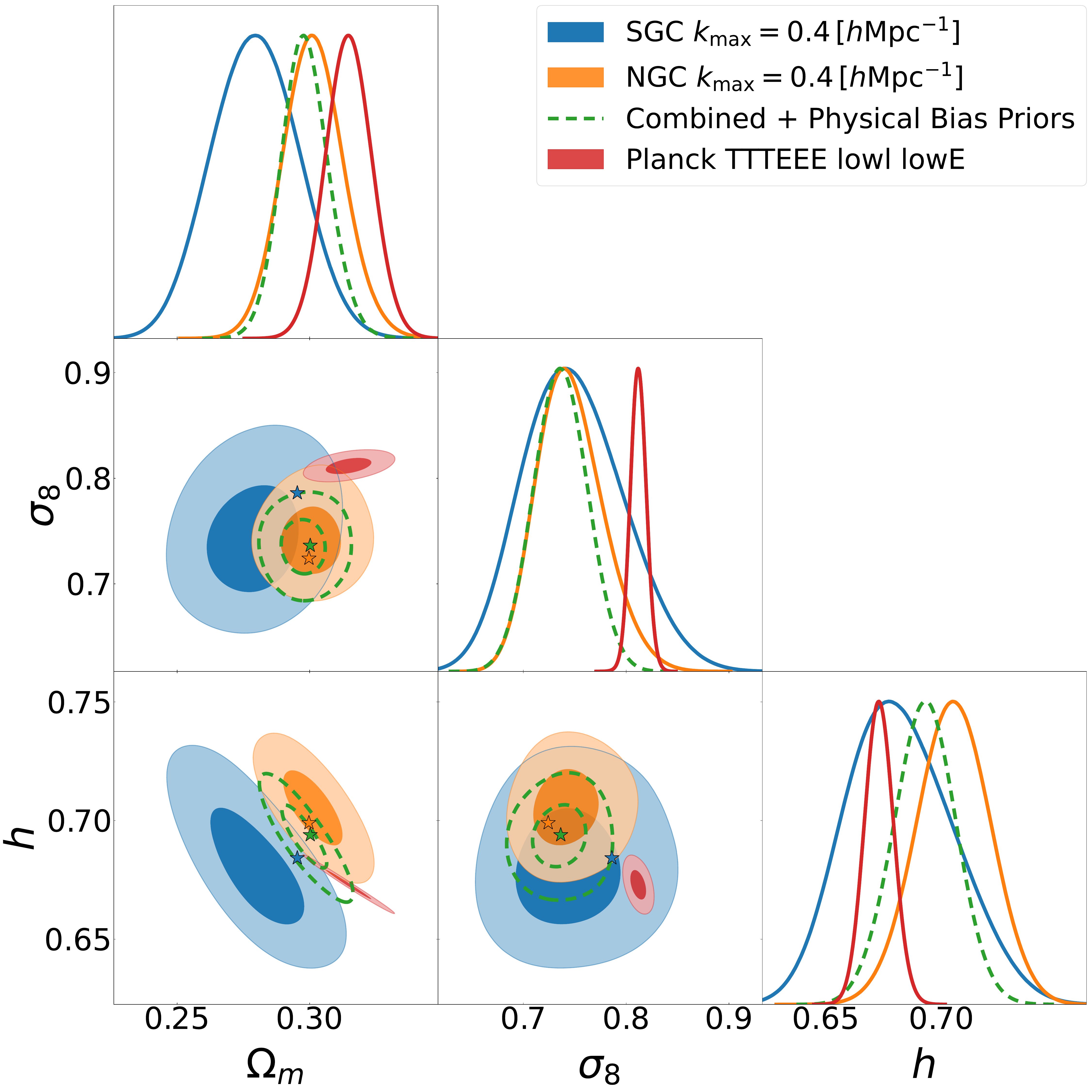}
   \caption{Constraints on the cosmological parameters $\om$, $\sigma_8$, and $h$ using the redshift-space power spectrum of galaxies in the SDSS-BOSS survey down to $k\simeq0.4\,\ihMpc$. We apply our methodology based on a hybrid bias expansion separately to two sub-regions of the survey, labelled as SGC and NGC (grey and red regions). In addition, as green contours, we report on the results of a joint analysis using relations between the higher-order bias parameters as measured from physical galaxy formation models. For comparison, we also display the results of the analysis of the {\it Planck\/} satellite data. In each case, the outer and inner contours indicate 68\% and 95\% credible intervals. Stars represent the best-fit values defined as the values that provide the maximum likelihood in the chains.}
    \label{fig:MoneyPlot}%
    \end{figure*}

\begin{figure*}
   \centering
   \includegraphics[width=\textwidth]{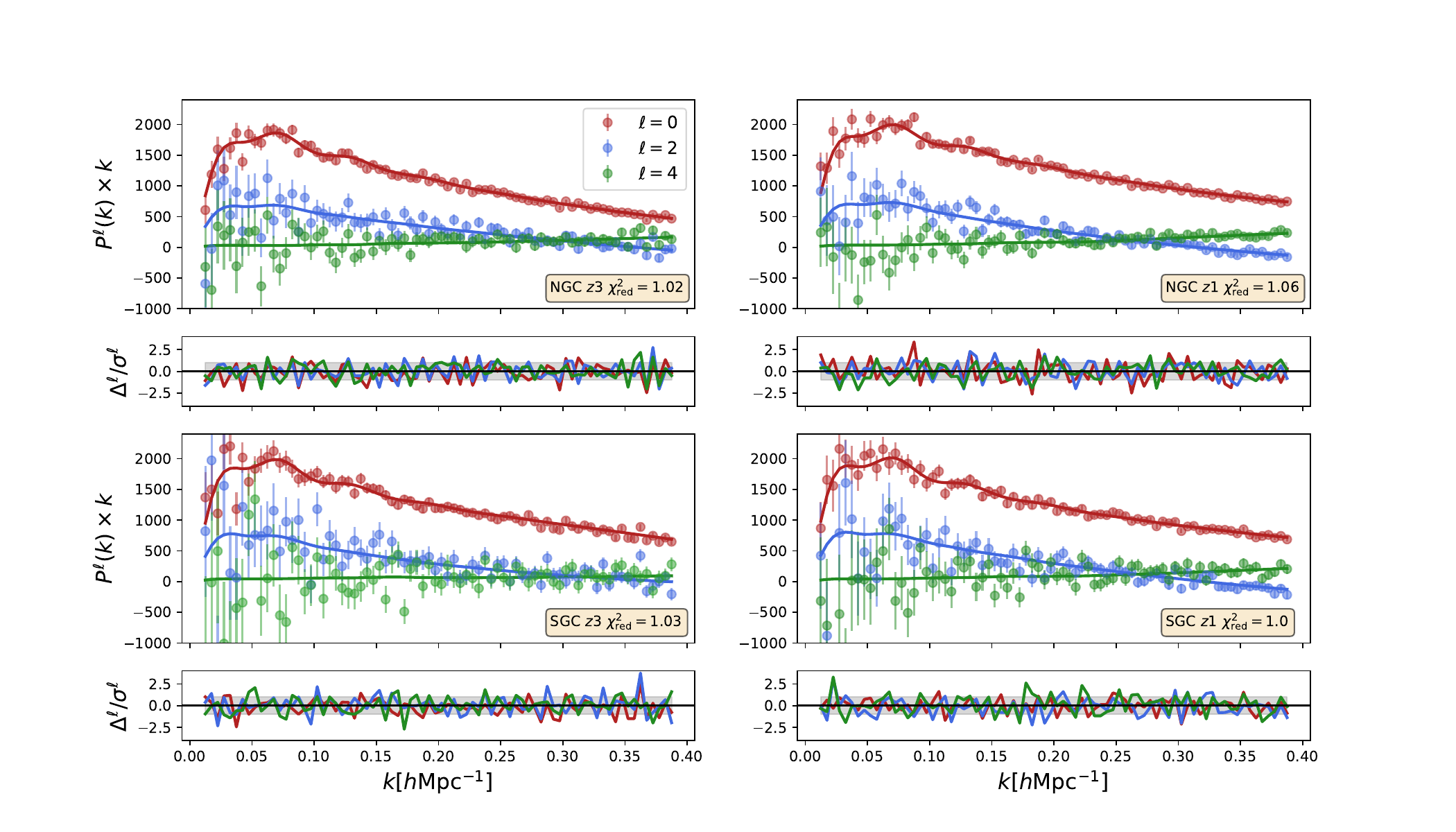}
   \caption{Comparison between the observed power spectrum monopole, quadrupole, and hexadecapole in SDSS-III BOSS (coloured symbols with errorbars), and the best-fit BACCO hybrid model (solid lines). Each panel shows one of the 4 subsamples of the BOSS data and the corresponding model prediction at the maximum of the posterior distribution function (see Fig.~\ref{fig:MoneyPlot}). The bottom plot of each panel displays the difference between model and data in units of the diagonal elements of the respective covariance matrix.
   }
    \label{fig:BestFits}%
\end{figure*}

After having validated our inference pipeline, we now present the key result of this work: parameter constraints from the BOSS DR12 dataset. As detailed in previous sections, our analysis relies on the BACCO hybrid Lagrangian bias expansion, which we apply to measurements of the monopole, quadrupole, and hexadecapole of the redshift space power spectrum in SDSS-III BOSS down to $0.4\,\ihMpc$. 

\subsection{Cosmological constraints from SDSS-BOSS}

In Fig.~\ref{fig:MoneyPlot} we present our constraints on the three main cosmological parameters we included in our analysis: $\om$, $\sigma_8$, and $h$. We display three cases. The blue and orange contours indicate constraints from the SGC and NGC data, respectively, where we jointly analysed multipoles from the low- and high-$z$ subsamples. The third case, indicated by dashed contours, corresponds to the combination of all 4 BOSS subsamples modelled adopting physical priors on the relationships of bias parameters (cf. \S\ref{sec:parameters}).\footnote{Note that combining NGC and SGC while varying all the bias parameters would require  37 free parameters (5 cosmological plus 32 nuisance). This is an extremely computationally expensive task for any nested sampling algorithm with the likelihood requirements of this analysis. Thus, since most of the constraining power in BOSS lies in NGC, we have refrained from analysing such case. We will explore methods to speed up the parameter estimation (such as the one presented in \citealt{Pellejero_2020}) in future works.}. For comparison, we also display the results obtained from the {\it Planck\/} satellite. The best-fit parameters as well as the 68\% confidence intervals are provided in Table~\ref{tab:parameters_table}. In Appendix A and Fig.~\ref{fig:FullPosterior} we show our constraints on the full parameter space.

Our fiducial NGC analysis yields a 3.6\% measurement on $\Omega_{\rm m}$, 4\% on $\sigma_8$, and 2\% on $h$. The accuracy on the derived parameter, $S_8 \equiv \sigma_8 \sqrt{\om/0.3}$, is $4.7\%$. In the case of the SGC dataset, we find $\sim30\%$ larger uncertainties, which is expected due to the volume $\sim3$ times smaller than NGC. 

Notice that the marginalised constraints from NGC and SGC are statistically consistent, with a mild disagreement in the $\Omega_m$-$h$ plane. However, there is a better agreement in terms of their best-fit parameters (shown as filled stars), which suggests the lack of significant sources of unaccounted systematic errors. In turn, the marginalised statistics might be affected (especially for SGC) by the boundaries of our priors and projection effects in the full parameter space.  

When considering the case of "Physical Bias Priors" (dashed contours), our estimates show very small shifts compared to those in our fiducial NGC analysis. Furthermore, this case shows a $\sim25\%$ decrease in uncertainties, mostly due to the combination of NGC and SGC. 

We have also analysed NGC and SGC separately including our bias priors. The resulting marginalised constraints are shown in grey in Fig.\ref{fig:ComparisonOthers}. We see that the constraints barely change in the case of NGC, but significantly improve in the case of SGC, due to the reduced freedom in the model. 


The consistency in our analyses indicates that the BOSS bias parameters are compatible with our assumed bias relations, as we will show explicitly below. Additionally, the increase in constraining power suggests that bias priors break internal model degeneracies, and are effective in reducing prior volume effects. 

Finally, we highlight that both of our models, and therefore the minimal $\Lambda$CDM, are a very good description of the BOSS data. We display an example of this in Fig.~\ref{fig:BestFits}, where we show the multipoles of each BOSS subsample together with the fiducial model evaluated at the best-fit parameter set obtained from the joint high-$z$ and low-$z$ NGC and SGC analysis. In the bottom panels, we display the difference in units of the diagonal elements of the covariance matrix. In each panel, we provide the value of $\chi^2_{\rm red} = \chi^2/{\rm d.o.f}$, where for simplicity we have estimated the number of degrees of freedom ${\rm d.o.f}$ as the number of datapoints minus the number of free parameters. The values are very consistent with 1, showing the hability of the model to fit the range of scales considered in this work. Although not shown, the $\chi_{\rm red}^2$ value increases slightly for the "Physical Bias Prior" case by approximately $\Delta\chi_{\rm red}^2 \sim 0.05$.


\subsection{Comparison to previous BOSS analyses}

\begin{figure*}
   \centering
   \includegraphics[width=0.8\textwidth]{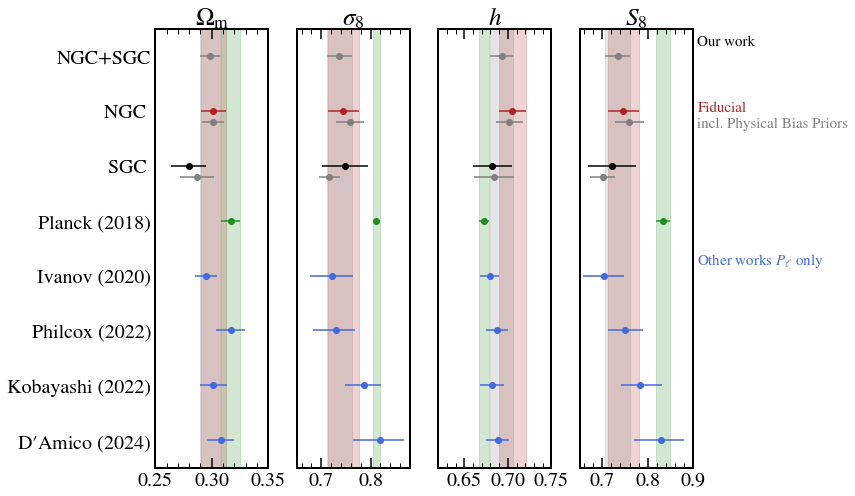}
   \caption{Comparison between our constraints on four cosmological parameters $\{\Omega_{\rm m}, \sigma_8, h, S_8\}$ with previous analyses of the SDSS-III/BOSS dataset. We display our constraints from using fiducial analysis pipeline and from assuming physical priors on the bias parameters as coloured and grey symbols, respectively. In all cases we employ the monolpole, quadrupole, and hexadecapole down to $\kmax\simeq0.4\,\ihMpc$. For comparison, we also display the results from {\it Planck\/} satellite as a green symbol and a shaded region. The measurements from other works have been taken from: \protect\cite{PlanckCosmo2020,Ivanov_2020,Philcox_2022,Kobayashi_2022,DAmico_2024}.}
 \label{fig:ComparisonOthers}%
\end{figure*}

The BOSS dataset is the largest publicly-available galaxy redshift survey, and has been previously analysed by a large number of works (see \citealt{Troster_2020} for one of the first examples on this ``full shape'' analysis). Fig.~\ref{fig:ComparisonOthers} provides a compilation of recent constraints on $\Omega_{\rm m}$, 
 $\sigma_8$, $h$, and $S_8$ using similar statistics to this work. 

More specifically, the same datavector we employ here was analysed by \cite{Ivanov_2020} and \cite{Philcox_2022}. They used EFT to model the power spectrum multipoles down to $\kmax \simeq 0.2\,\ihMpc$. \cite{DAmico_2024} used a similar model, but on the windowed BOSS data \citep{Beutler_2021} up to $\kmax=0.23\,\ihMpc$. Finally, \cite{Kobayashi_2022} analysed the data with an HOD emulator up to $\kmax = 0.25\ihMpc$, also with the windowed version of the datavector. As in our analyses, these works adopt informative priors on $n_s$ based on {\it Planck\/} and on $\Omega_bh^2$ based on BBN \citep{Cooke_2018}. 

Overall, we find that  our constraints agree statistically with those obtained in previous works. However, our analysis delivers $\sigma_8$ measurements which are $30\%$ and $40\%$ more accurate in the fiducial NGC and physical bias prior cases, respectively. There is, however, considerably scatter among the measurements of $\sigma_8$, as seen in the second column of Fig.~\ref{fig:ComparisonOthers}. This could suggest that the specific choices of each analysis plays an important role. 

Specifically, previous works based on EFTofLSS \citep{Carrasco_2014} used $\kmax \simeq 0.2\,\ihMpc$, since the model accuracy drops rapidly on smaller scales. In contrast, our model has been shown to work well on mocks down to scales of $k\simeq 0.6\ihMpc$. There are two possible differences between EFT approaches and our approach. On one side, we are pushing our study to smaller scales which potentially carry information. As we will see in \S\ref{sec:kmax}, the increase on constraining power with scale in our fiducial case is negligible, therefore making this explanation highly unlikely. On the other side, our model does not need the use of ``counterterms'', since all the non-linear evolution is encoded in the N-body displacement field\footnote{Note that, even though we don't need such counterterms for the real space theory, the FoG parameters can be regarded as counterterms partially accounting for the RSD non-linearities.}. This changes the shape of the parameter space, changing the degeneracies between cosmological and nuisance parameters. 

The derived structure parameter $S_8$ is also consistent among different analyses. The biggest disagreement is with \cite{DAmico_2024}, who reported a systematically higher value for $S_8$ than every other BOSS analysis. This result, however,  includes a correction for prior volume effects, which makes it difficult to compare against other measurements. In contrast, our constraints at $\kmax\simeq0.4\ihMpc$ do not show evidence of strong prior volume effects, as we will discuss in \S\ref{sec:prior_volume}.

The BOSS multipoles were analysed jointly with additional summary statistics by \cite{Philcox_2022}, \cite{Chen_2022}, \cite{Ivanov_2023}, and \cite{DAmico_2024}, who included combinations of the galaxy bispectrum monopole, quadrupole, an analogue to the real-space power spectrum, and the post-reconstruction BAO feature. Consistent with these works, our results are in agreement in terms of the marginalized statistics and remain among the most precise determinations of $\sigma_8$ despite using a reduced data vector. For instance, the measurements of \cite{Ivanov_2023}, which incorporated all of the aforementioned statistics, yield $H_0=68.2\pm 0.8$ km s$^{-1}$Mpc$^{-1}$, $\Omega_{\rm m} = 0.33\pm 0.01$, $\sigma_8=0.736\pm0.033$, and $S_8= 0.77\pm0.04$. Compared to our results, their analysis shows 38\% stronger constraints for $H_0$, arguably due to the post-reconstruction BAO, but slightly looser constraints for $\om$ and $\sigma_8$ (14\% and 24\%, respectively). 

Another interesting comparison is with the results of \cite{Kobayashi_2022}. Using an emulator for the redshift-space power spectrum of halos combined with the HOD model \citep{Kobayashi_2020}, they were able to analyse scales up to $\kmax \approx 0.3\ihMpc$. Although our approach to include RSD in the model differs, it is similarly based on halo velocities. Despite differences in our modelling strategies, we find consistent results. Notably, our method yields a slight increase in accuracy for $\sigma_8$, achieving 4\% accuracy compared to their 5\%. This is a minor difference given the significantly different analysis methods. Nonetheless, both studies highlight the potential of using $N$-body emulator-based models to extract cosmological information.

Finally, it is noteworthy that our constraints on $h$ and $\Omega_{\rm m}$ align remarkably well with those obtained by the latest BAO results from the DESI collaboration. The values presented in \cite{DESIBAOCosmo} are $\Omega_{\rm m}=0.295 \pm 0.015$ for DESI alone and $h=0.6929 \pm 0.0087$ when combined with $r_d$ from the CMB. This consistency follows the trend toward lower values of $\Omega_{\rm m}$ and higher values of $h$ observed in Fig. \ref{fig:MoneyPlot}.



\subsection{Constraints on galaxy bias parameters}

\begin{figure}
\centering
   \includegraphics[width=\columnwidth]{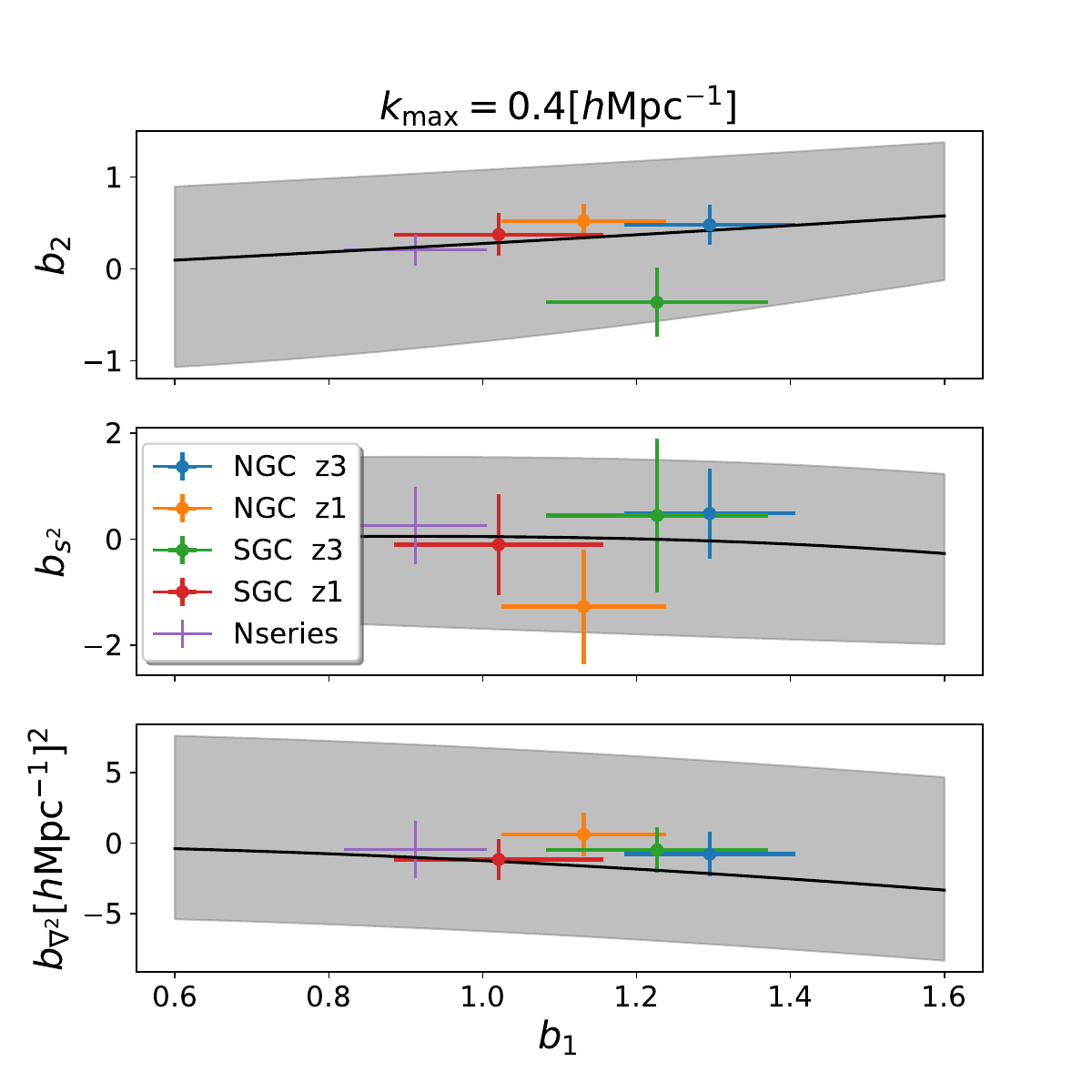}
   \caption{Constraints on the Lagrangian bias parameters, $b_1$, $b_2$, $b_s^2$, and $b_{\nabla}^2$ from the analysis of redshift-space multipoles down to $k\simeq 0.4\,\ihMpc$.  We display our results from each of the 4 SDSS-III/BOSS subsamples (cf. \S\ref{sec:data}), and for the suite of \nseries\ mocks, as indicated by the legend. The shaded regions and black line correspond to the $3\sigma$ and mean relationship provided by \protect\cite{Zennaro_2022} from the analysis of thousands of galaxy formation simulations.}
 \label{fig:Coevolution}%
\end{figure}

The Lagrangian bias parameters measure how, on average, the number of galaxies is modified by large-scale features of the Lagrangian density field. For instance, $b_1$ measures how galaxies respond to large-scale changes in density, whereas $b_{s^2}$ does so for changes in the tidal field. In this sense, the bias parameters have a clear physical meaning, and it is therefore interesting to compare their values in observations to theoretical expectations, as this is a direct test of galaxy formation models. 

In Fig.~\ref{fig:Coevolution} we show our constraints on the Lagrangian bias parameters from the analysis of the redshift-space multipoles, using $\kmax=0.4\ihMpc$. Note that we display our results in terms of the linear bias, $b_1$, since the absolute value of the bias parameters is much more sensitive to cosmology and details of the galaxy population than their relation to $b_1$ is. We display results from the analysis of BOSS data and of the \nseries\ mocks. For comparison, we also display the relationship measured by \cite{Zennaro_2022} from a large suite  SHAMe galaxy mocks, exploring widely different galaxy formation parameters, redshifts, cosmologies, and number densities, including one sample that resembles the clustering of galaxies stellar-mass selected galaxies in the IllustrisTNG simulation with the same number density as BOSS CMASS sample.

We find that the recovered bias parameters agree with the \cite{Zennaro_2022} relationship within the expected uncertainty. In particular, the {\nseries} sample is within $0.4\sigma$ of the mean predicted value, albeit the $b_1$ value being systematically smaller than the BOSS high-$z$ data (this is in part because the \nseries\ simulations adopted a 15\% higher $\sigma_8$ than our BOSS best-fit value). The bias parameters in BOSS show more scatter than in \nseries\ (since for the latter we employ the mean of 84 mocks) but always lie within the predicted expected region and very close to the mean relation (shown by solid black lines). 

The fact that the prediction roughly agrees with the measured values indicates that, even though the bias parameters might absorb some of the observational systematic not taken into account in the observational weights, they seem to retain their physical meaning as response functions to the large scale density field.

The only subsample that shows somewhat of a tension with the theoretical expectations is the SGC high-$z$ sample. In fact, this sample not only deviates from the mean $b_1-b_2$ and $b_1-b_{s^2}$ relationships, but the value of $b_1$ is also different from that of the NGC high-$z$ despite both samples containing similar types of galaxies. Although this tension is not strong from a statistical point of view, we note that when we adopt bias priors in the SGC analysis, the derived value is $b_1=1.49$, in much better agreement with the NGC counterpart.


\section{Robustness tests}
\label{sec:robustness}

In this section, we explore the robustness of our results to various analysis choices. Specifically, we quantify the sensitivity of our cosmological constraints to $\kmax$, the importance of the hexadecapole, and the impact of prior volume effects.

\subsection{The impact of $k_{\rm max}$ }
\label{sec:kmax}

\begin{figure}
   \includegraphics[width=\columnwidth]{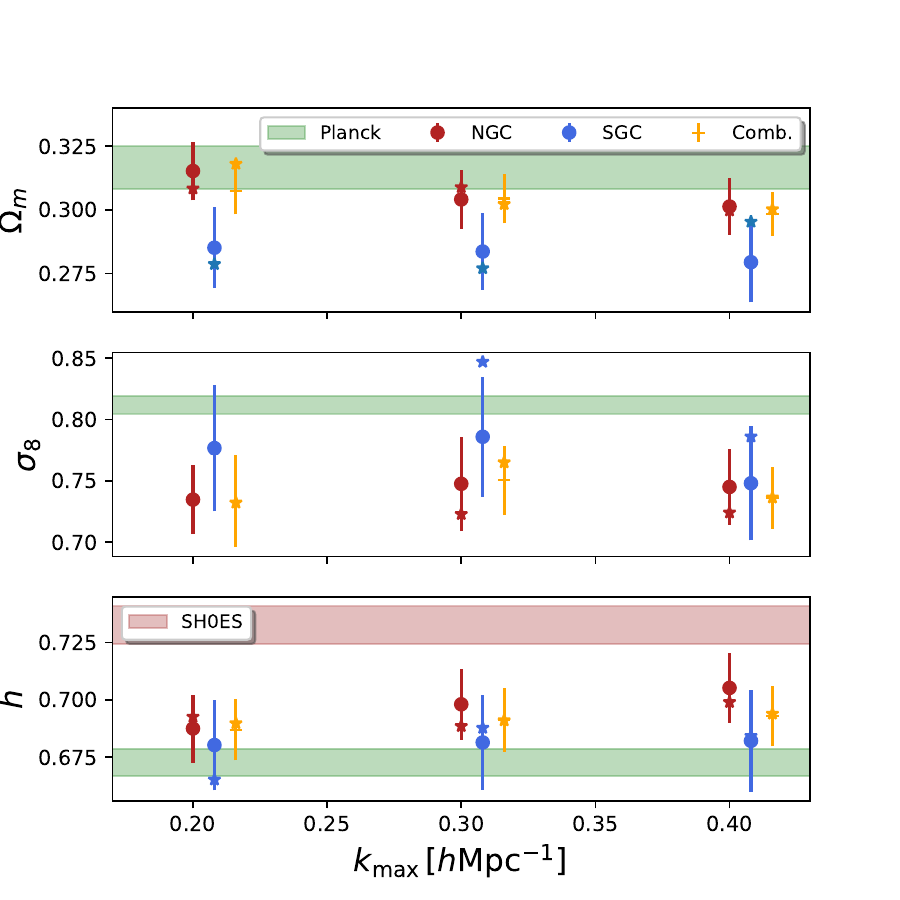}
   \caption{Consistency of our cosmological constraints as a function of the maximum wavenumber included in the analysis, $\kmax$. The plot displays our measurements on $\Omega_m$, $\sigma_8$, and $h$ with their corresponding $1\sigma$ uncertainties. We find no evidence of tensions between the inferred parameters as a function as we include small scales. The star symbols represent the best-fit values. They show a greater level of consistency in both $\Omega_{\rm m}$ and $h$ for our fiducial results.}
    \label{fig:ScaleConsistency}%
\end{figure}

In Fig.~\ref{fig:ScaleConsistency} we compare the marginalised constraints on $\Omega_{\rm m}$, $h$, and $\sigma_8$ as a function of the maximum scale included in the analysis, $\kmax$. We display our fiducial analysis carried out independently on NGC and SGC data, and of their joint analysis in our ``physical bias priors'' case.

Firstly, we see that our constraints remain remarkably stable from $\kmax=0.2$ to $0.4\,\ihMpc$ (note we do not consider $\kmax \leq 0.1\,\ihMpc$ as these constraints are heavily affected by the limits adopted in our emulator). In almost all cases, the parameter shifts are below $0.4\sigma$, with the exception of $h$ in the NGC case, which shifts by $0.7\sigma$. However, $h$ is extremely stable when we adopt bias priors, which suggest that, at least part of this can be explained by internal model degeneracies at $\kmax=0.2\,\ihMpc$. This is also evident from the comparison of the best fit values, shown as star symbols, which display a higher level of consistency than the marginalised statistics in the fiducial case. This indicates that the best fit obtained at $\kmax=0.4,\ihMpc$ also provides a good fit at $0.2,\ihMpc$. We verified this and found a value of $\chi^2_{\rm NGC,red}(\kmax=0.2,\ihMpc) \approx 1.04$.

Secondly, we note that the accuracy on cosmological parameters does not improve significantly with $\kmax$ in our fiducial NGC and SGC analyses. This is consistent with our results on the \nseries\ mocks (cf. \S\ref{sec:nseries}) and with the findings of \cite{PellejeroIbanez_2024}, where we used the BACCO hybrid bias emulator to analyse a mock catalogue resembling BOSS. Due to the limited volume and number density of BOSS galaxies, the additional small-scale Fourier modes mostly constrain the nuisance parameters of the model. For instance, the accuracy with which the bias parameters are measured increases by a factor of 30\% from $\kmax=0.2$ to $0.4$ (mainly in $b_2$ and $b_{s^2}$), and by 90\% in the case of the parameters modelling small-scale velocities ($\lambda_{\rm FoG}$ and $f_{\rm sat}$).

Given that small scales in BOSS mostly constrain nuisance parameters of our model, it is not surprising that in the ``Physical Bias Priors'' case we detect a mild but steady increment in cosmological information with $\kmax$. For instance, the estimate of $\sigma_8$ improves by 32\% compared to $\kmax=0.2\ihMpc$. However, the gains are still moderate considering the enormous increase in the number of modes. This is because of the additional model freedom provided by the nuisance parameters describing redshift-space distortions and the stochasticity in the matter-halo connection. This highlights the potential gains of investigating and placing robust priors on these parameters, in an analogous way to the bias parameters.

\subsection{Prior volume effects}
\label{sec:prior_volume}
\begin{figure}
   \includegraphics[width=\columnwidth]{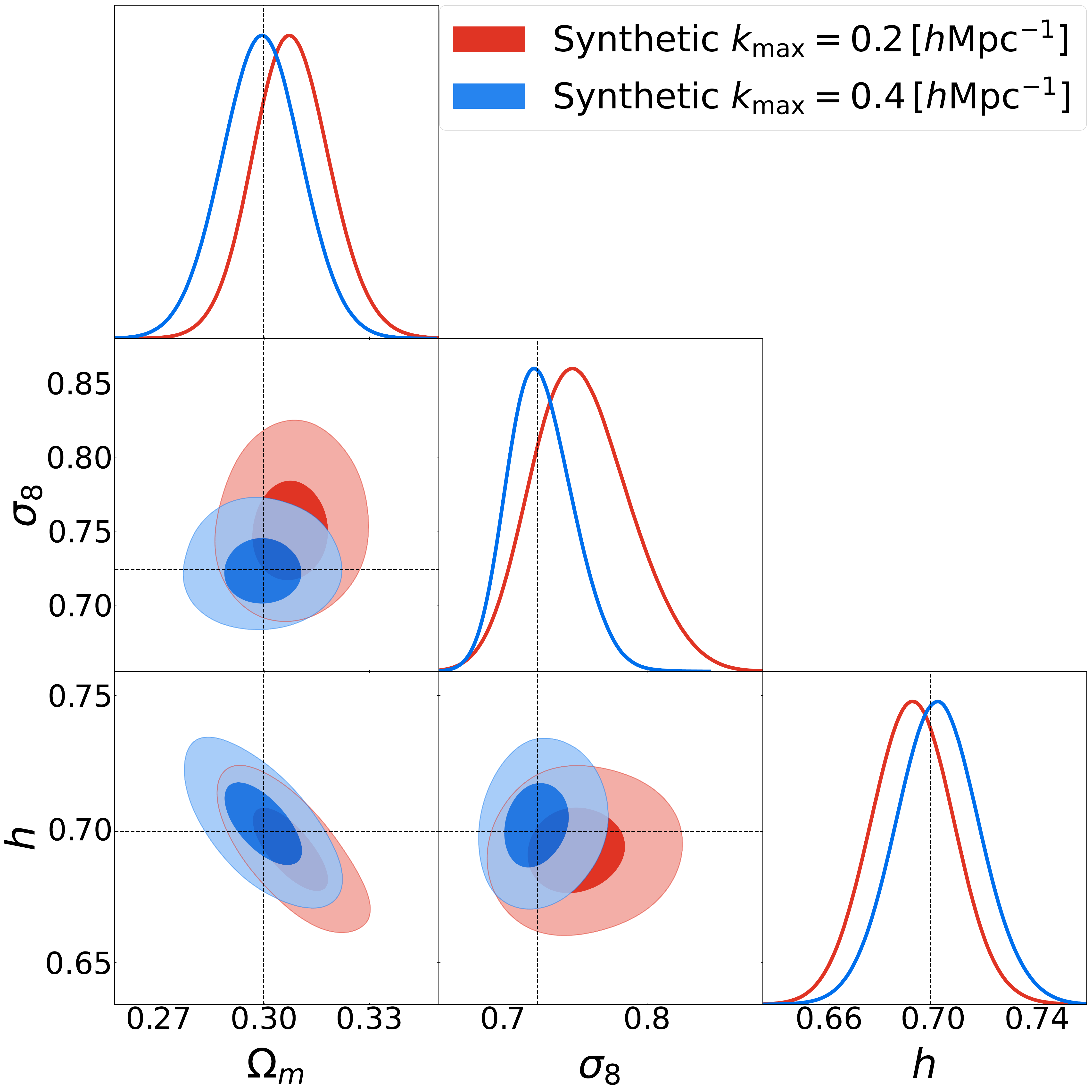}
\caption{Test on the importance of prior volume effects for our results. We display the $1$ and $2\sigma$ confidence intervals on $\om$, $\sigma_8$, and $h$ obtained using our methodology on a synthetic datavector down to $\kmax=0.2\,\ihMpc$ or $0.4\,\ihMpc$. The datavector was generated by evaluating our model at the best-fit values of our fiducial BOSS analysis (indicated by the dotted lines). The data vector is then noiseless and represents the mean of all possible realisations of the theory given by the covariance defined by the \texttt{Patchy} mocks. Therefore, any deviation from the input values indicates that the marginalised cosmological constraints are affected by unconstrained parameters of the model.}
    \label{fig:prior_volume}%
\end{figure}

Due to the complexity of our model, there are several nuisance parameters that could be loosely constrained, especially when considering low values of $\kmax$. In such cases, the marginalised constraints on cosmological parameters might suffer from "prior-volume effects", that is, the estimated posterior becomes sensitive to the prior distribution, boundaries adopted, and projections from a high dimensional parameter space. In the context of EFTofLSS, \cite{Carrilho_2023} and \cite{DAmico_2024} have cautioned that the analysis of BOSS might suffer from such projection effects. Although the relevance of priors is a key aspect of Bayesian statistics, it is important to identify the existence of such effects, otherwise it might mislead the interpretation of results.

In Fig.~\ref{fig:prior_volume}, we present a quantification of prior-volume effects in our results. Specifically, we present the analysis of a datavector generated by evaluating our model with the best-fit values of the combined BOSS-NGC sample (see Table~\ref{tab:parameters_table}). The data vector is then noiseless and represents the mean of all possible realisations of the theory as defined by the covariance from the \texttt{Patchy} mocks. Naturally, we expect the best-fit parameters to exactly coincide with the input values (indicated by dotted lines). Therefore, any deviation between the input and the marginalised values can be attributed to the effect of priors.

We can see that projection effects are almost negligible in our fiducial analysis of NGC, implying our current priors are uninformative. This also eliminates the need for correcting our results in a manner explored by \cite{DAmico_2024}. In contrast, when the model is used only up to $\kmax=0.2~\ihMpc$, the mean of the marginalised posterior on $\sigma_8$ and $\Omega_{\rm m}$ overestimate the true values by almost $1\sigma$, as a consequence of poorly constrained nuisance parameters. Note that this does not entirely demonstrate that we are unaffected by these effects; rather, it shows that we are not dominated by them at the maximum likelihood values of our parameter space. Different regions will be affected differently by prior-volume effects. To completely rule out being dominated by them, a profile likelihood analysis would be required. However, this is computationally expensive, and we leave such a test for future work.



\subsection{The impact of the hexadecapole}
\label{sec:hexa}

    \begin{figure}
   \includegraphics[width=\columnwidth]{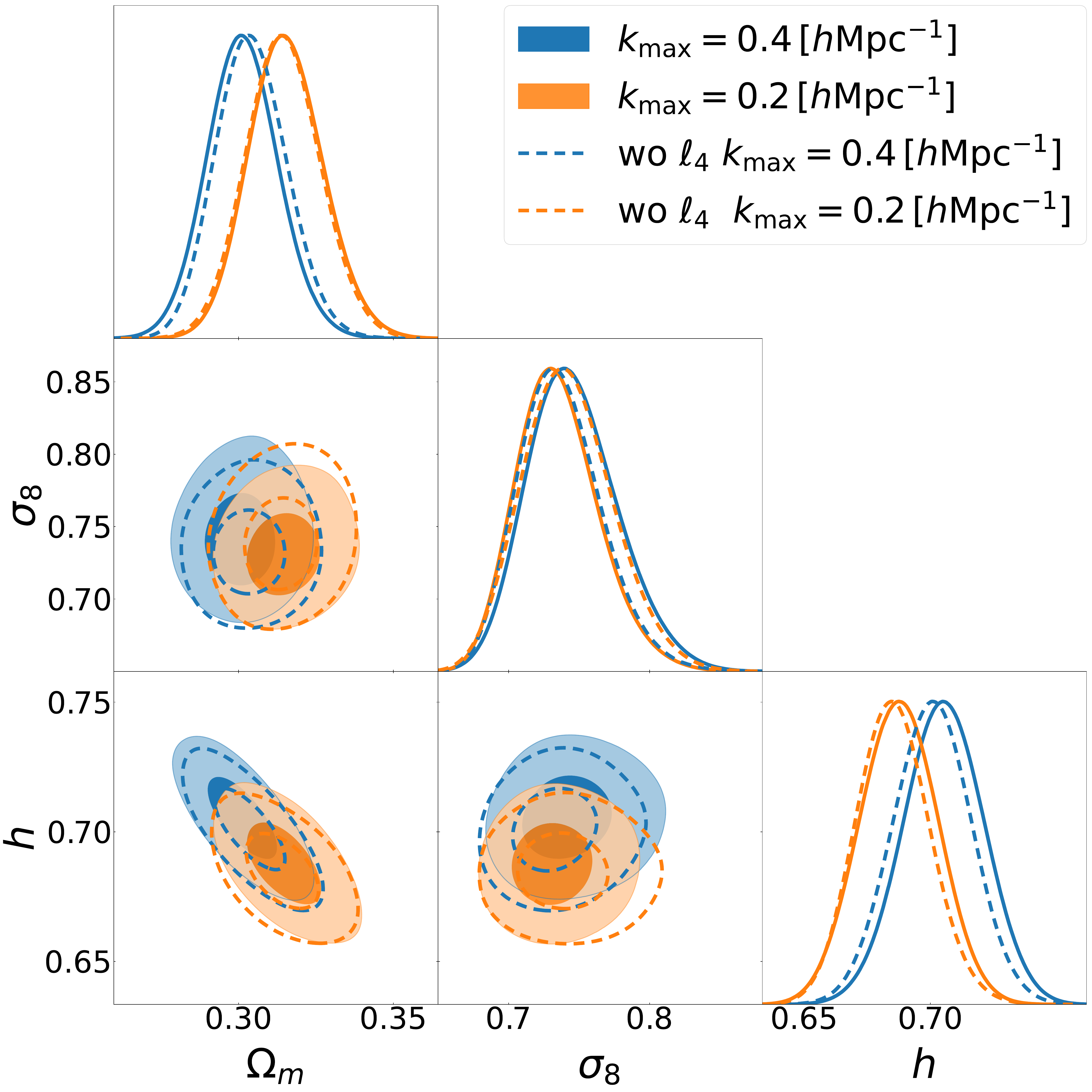}
   \caption{Robustness of our cosmological constraints to the inclusion of the hexadecapole of the redshift-space BOSS-NGC power spectrum. We display as blue contours our fiducial $\kmax\simeq0.4\,\ihMpc$ analysis of the NGC sample, whereas in orange we show the same analysis but at $\kmax\simeq0.2\,\ihMpc$. The dashed lines show the results excluding the hexadecaople from our data vector, as indicated by the legend.}
    \label{fig:hexaeffect}%
    \end{figure}

The hexadecapole of the galaxy power spectrum is the most sensitive to the line-of-sight correlations, and hence to the physics causing small-scale velocities. In addition, it is the noisiest multipole predicted by our emulator, owing to the finite volume of the simulations employed. Additionally, in \cite{Beyond-2pt_2024}, we showed that the uncertainties in the emulation of the hexadecapole could introduce small but significant biases in our estimation of $\sigma_8$. 

To explore the robustness of our results to the modelling of the hexadecapole, we re-analyzed the BOSS-NGC data, but only considering the monopole and quadrupole. In Fig.~\ref{fig:hexaeffect} we present our results for $\kmax=0.2$ and $0.4\,\ihMpc$. In both cases, we see that the hexadecapole shifts the inferred parameter values by less than $0.2\sigma$ -- a change that is statistically insignificant compared to other sources of noise. This is consistent with our expectations from the analysis of the \nseries\ mocks, which showed no significant parameter biases, and with \cite{PellejeroIbanez2023} where we showed that the emulator noise is below the statistical uncertainty of BOSS data. 

Finally, it is worth noting that although the hexadecapole has a negligible impact on the precision when $\kmax=0.4\,\ihMpc$, it improves the accuracy in the $\kmax=0.2\ihMpc$ case, where, for instance, the uncertainty in $\sigma_8$ decreases by 10\%. Additionally, when excluding the hexadecapole, the cosmological constraints consistently improve with $\kmax$, unlike in our fiducial analysis (cf. \S\ref{sec:kmax}). This indicates that at least part of the information encoded on small scales is already captured by the hexadecapole on large scales.



\section{Discussion}
\label{sec:discussion}

\subsection{The $S_8$ tension}

\begin{figure}
   \centering
   \includegraphics[width=1\columnwidth]{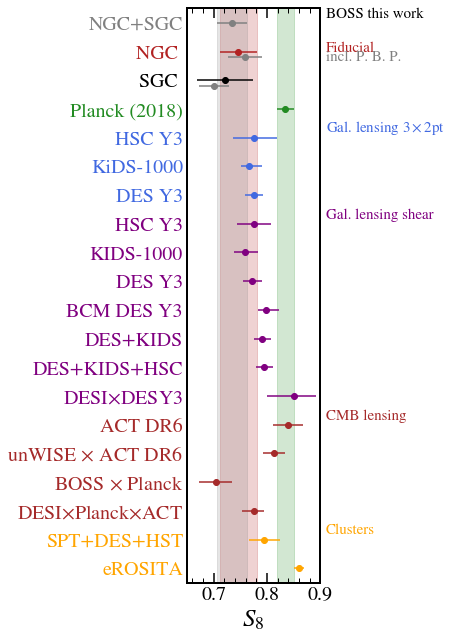}
   \caption{Constraints on the ``structure parameter'' $S_8$ from our fiducial and "Physical Bias Priors" analyses of SDSS-III/BOSS and from other large-scale structure probes. For comparison, we show the value inferred by {\it Planck\/} as a green symbol and shaded region. The data shown in this plot comes from the works by \protect\cite{PlanckCosmo2020,Garcia-Garcia2024,ACTCosmo2024,HSCY3_2023,Heymans2021,DESY32022,Arico2023,DESKIDS_2023,unWISEACT_2024,SPTDESHST2024,eROSITA_2024,Sugiyama_2023,Amon_2022,Asgari_2021,Chen_2024,Sailer_2024,ChenDESIxDES_2024}. }
    \label{fig:ComparisonOthersS8}
\end{figure}

It has become increasingly common to characterise the amplitude of matter fluctuations in the Universe via the ``structure parameter'' $S_8$. The motivation behind this is that weak gravitational lensing directly measures this parameter combination, which has influenced its use in other low-$z$ cosmic probes. Comparing direct measurements of $S_8$ at low redshift, with the expectations of high-$z$ measurements offers a direct way to test whether the growth of structure is compatible with the predictions of the $\Lambda$CDM model.

With the arrival of Stage-III lensing surveys, a tension started to emerge between low-$z$ measurements of $S_8$ and the expectation from analyses of the {\it Planck\/} satellite. For instance, the analysis of shear correlations as measured by DES-Y3, HSC-Y3, and KIDS1000, all reported a tension with {\it Planck\/}, ranging between $2$ and $3\sigma$ \citep{Heymans2021,Secco:2022,HSCY3_2023,Asgari_2021}. Similarly, the joint analysis of lensing with photometric galaxy clustering (a.k.a. 3$\times$2pt analyses), strengthened this conclusion, typically reporting a slightly larger tension \citep{Heymans2021,DESY32022}. One of the most precise measurements of $S_8$ was provided by \cite{Garcia-Garcia2021}, who combined multiple surveys to find a value $3.4\sigma$ smaller than in {\it Planck\/}. Similarly, analysis of the cross-correlation between galaxies and CMB lensing, and of the abundance of clusters, as well as the analyses of BOSS clustering, all reported values systematically below {\it Planck\/} \citep{Bocquet:2019,Krolewski2021,White2022,Ivanov_2020}. Additionally, \cite{Amon2023} suggested that a low $S_8$ value could be the solution for the ``lensing-is-low'' problem \citep{Leauthaud2017}. The state was summarised in a 2021 Snowmass report \citep{DiValentino2021}.

More recently, the situation has started to change. As emphasised in an independent reanalysis of DES-Y3, \cite{Arico2023} cautioned that current weak lensing constraints were sensitive to the model for the nonlinear power spectrum, the treatment of intrinsic alignments, and the assumptions regarding baryonic physics. Explicitly, using the BACCO emulators \citep{Arico_2021,baccoemu2023}, \cite{Arico2023} found a value for $S_8$ compatible with {\it Planck\/} at the $1.4\sigma$ level. This was also the conclusion of a subsequent official reanalysis of the DES and KIDS surveys \citep{DESHSC2023}. Additionally, an important point to note is that most of the constraining power of weak lensing arises from nonlinear scales. Therefore, as highlighted by \cite{Amon2022}, agreement between lensing and {\it Planck\/} can be obtained if baryonic physics is stronger than the level that is typically assumed in hydrodynamical simulations.

An alternative way of measuring gravitational lensing is through its effect on the CMB: this signal is well consistent with the prediction of the {\it Planck\/} cosmology \citep{PlanckCosmo2020, ACTCosmo2024}. However, the total CMB lensing signal comes from much higher redshifts than those probed by weak galaxy shear. A tomographic study that isolated the $z<0.8$ part of the CMB lensing signal found a result lower than {\it Planck\/} \citep{Hang2021}, arguing that this could be consistent with both the low $S_8$ results and the good agreement between total CMB lensing and {\it Planck\/} if the tension was mainly in the direction of lower $\Omega_m$.


Finally, the most recent analyses of the abundance of clusters detected by the SPT survey \citep{SPTDESHST2024} and of the cross correlation of {\it Planck\/}-CMB lensing with galaxies in the WISE survey \citep{unWISEACT_2024} both reported milder tensions with {\it Planck\/}. The eROSITA cluster abundances \citep{eROSITA_2024} even yielded values of $S_8$ slightly above that of {\it Planck\/}, with a nominal accuracy that even exceeds that of {\it Planck\/}. Finally, \cite{Chaves-Montero2023} and \cite{Contreras2023b,Contreras2023} showed that an apparent ``lensing-is-low'' tension  appear due to the limitations of simplistic models for the galaxy-halo connection.

In Fig.~\ref{fig:ComparisonOthersS8} we show a compilation of recent constraints on $S_8$, compared with our BOSS results (shown in the first three rows). We can see that our results are statistically compatible with most of the recent low-$z$ cosmic probes, perhaps with the exception of eROSITA. Moreover, our reported $S_8$ value is $2-3\sigma$ lower than {\it Planck\/} with a mean value towards the lower end of values in the literature, although still compatible with other full shape analyses such as \cite{Ivanov_2023}. Overall, the $S_8$ tension is statistically mild: given the scatter between the non-{\it Planck\/} measurements, it is hard to make an unambiguous case that there is an inconsistency that requires systematics.
Nevertheless, almost all alternative measurements argue that $S_8$ 
should be below the central {\it Planck\/} value, and so at a minimum there is a good case that the {\it Planck\/} value has fluctuated high.


In the next subsections we will explore the $S_8$ tension further, first in terms on an alternative parameterisation for the amplitude of structure fluctuations, $S_{12}$, and then on the feasibility of the {\it Planck\/} cosmology to explain BOSS data.

\subsection{$S_{12}$ and the $H_0$ tension}
\label{sec:s12}

\begin{figure}
\includegraphics[width=\columnwidth]{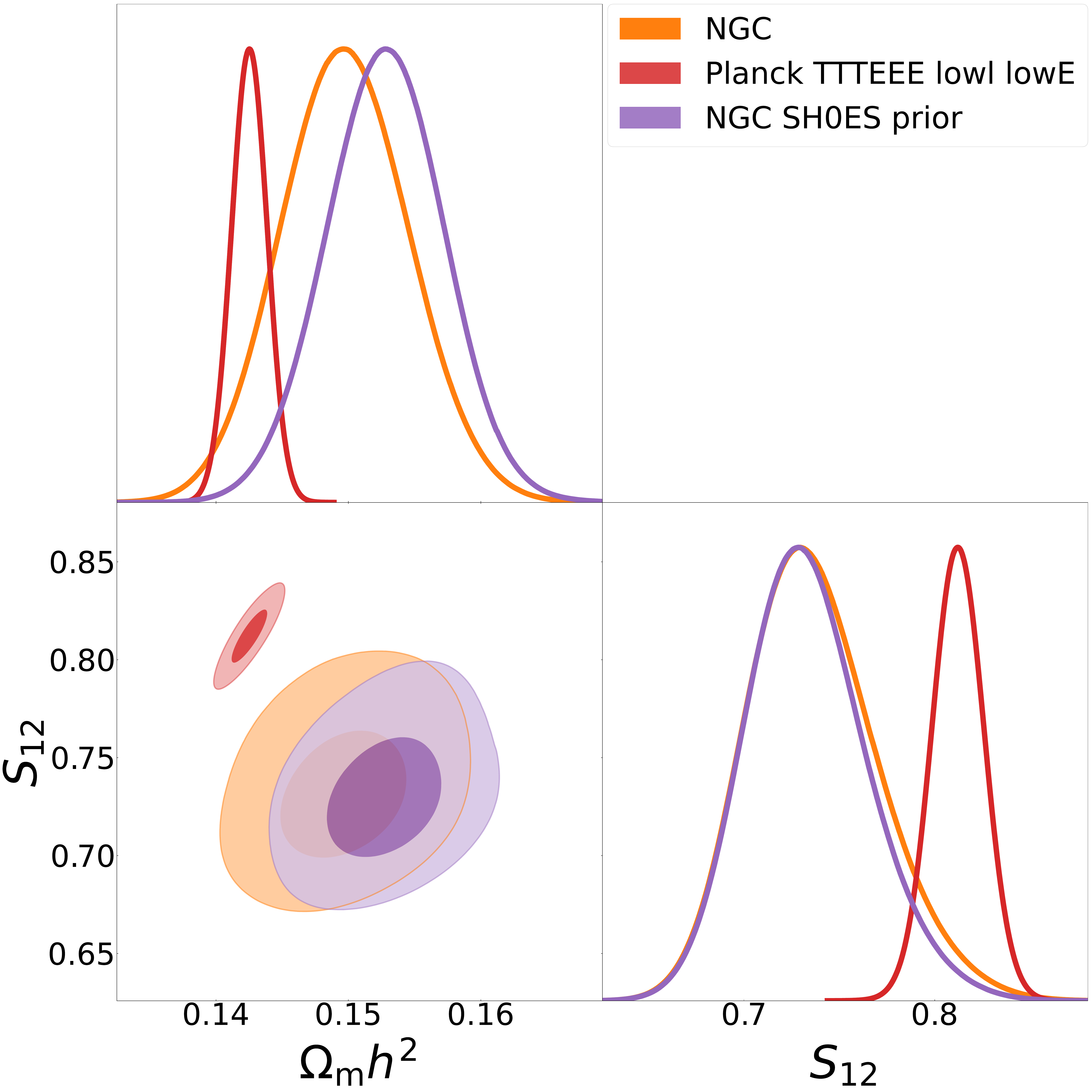}
\caption{Constraints on the cosmological parameters $\omega_m$ and $S_{12}$ obtained from our fiducial analysis of BOSS-NGC up to scales $\kmax\simeq 0.4\ihMpc$. Orange contours show our results using uninformative priors on the Hubble parameter, $H_0$, whereas purple contours impose a prior from measurements in the local Universe from SH0Es.}
\label{fig:S12}%
\end{figure}

In light of the tension between our best-fit parameters with {\it Planck\/}, in this section we explore a different parameterisation and the role of an informative prior on $H_0$.

Motivated by \cite{Sanchez2020}, we consider the parameters $\omega_m \equiv \Omega_m h^2$ and $S_{12} = \sigma_{12} (\omega_m/0.14)^{0.4}$, where $\sigma_{12}$ is the rms linear fluctuations in $12$ Mpc spheres. The advantages of this parameterisation are that the value of $h$ does not enter in the definition of $S_{12}$ and that there is a separation into parameters that affect the shape and those that affect the amplitude of the linear power spectrum. Furthermore, as discussed in \cite{Sanchez2020} and \cite{Garcia-Garcia2024}, weak lensing constrains the overall amplitude of fluctuations which makes it sensitive to the adopted prior on $h$. In fact, \cite{Garcia-Garcia2024} showed that the tension in $S_8$ between weak lensing and {\it Planck\/} does not appear either in $S_{12}$ or when a prior on $H_0$ based on SH0ES data is employed, suggesting that the current $S_8$ tension might be another face of the $H_0$ tension between {\it Planck\/} and SH0Es.  

In Fig.~\ref{fig:S12} we display our constraints on $\omega_m$ and $S_{12}$ from our fiducial NGC analysis. Unlike in weak lensing analyses, adopting  this parameterisation does not relax the tension between BOSS and {\it Planck\/}. Additionally, imposing a prior on $H_0$ from local measurements from SH0ES ($h = 0.7304 \pm 0.0104$, \citealt{SH0ES_2022}) does not alleviate the tension with {\it Planck\/} -- in fact, the discrepancy in $S_{12}$ increases slightly. The main reason behind these results is that, unlike weak lensing analyses, full-shape clustering already constrains $H_0$ very well ($\sim2\%$, similar to the accuracy of SH0ES), which implies an almost unique relationship between $S_{12}$ and $S_8$. Hence, a prior on $H_0$ will only have a moderate effect on the constraints from redshift-space clustering.

\subsection{Comparison with {\it Planck\/} cosmology}
\label{sec:planck}

\begin{figure}
\includegraphics[width=\columnwidth]{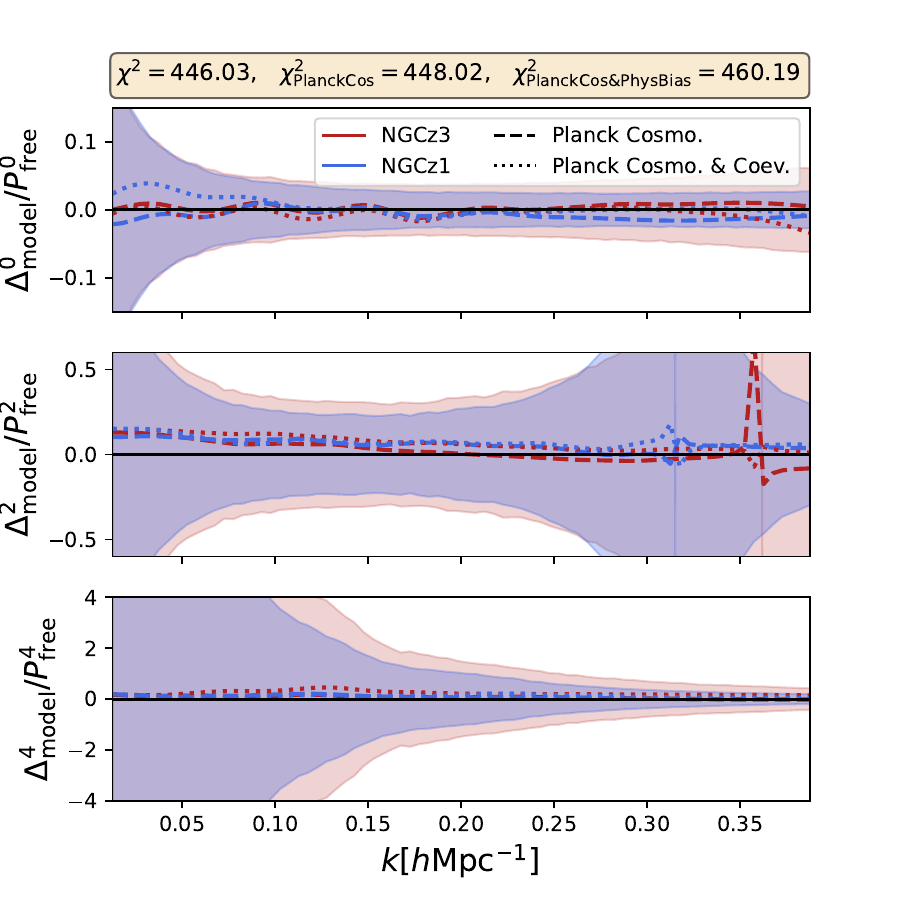}
\caption{Fractional difference between best-fit models to BOSS-NGC from our fiducial analysis and when we fix the cosmological parameters to {\it Planck\/} values. The shaded regions correspond to the $1\sigma$ uncertainty in the measured multipoles, as estimated using the \texttt{Patchy} mocks. The corresponding values of the $\chi^2$ are shown in the legend. }
\label{fig:PlanckFix}%
\end{figure}


In \S\ref{sec:inference} we showed $\Lambda$CDM to be an excellent description of the data, however, as can be seen in Figs.~\ref{fig:MoneyPlot} and \ref{fig:ComparisonOthersS8}, the parameter constraints differ with those obtained from the {\it Planck\/} satellite.

We quantify the tension between BOSS and {\it Planck\/} in the full cosmological space as
\begin{equation}
    \chi^2 = (\mu_{\it Planck\/} - \mu_{\rm BOSS})(C_{\it Planck\/} + C_{\rm BOSS})^{-1} (\mu_{\it Planck\/} - \mu_{\rm BOSS})^T
\end{equation}

\noindent where $\mu \equiv\{\Omega_m, \sigma_8, h\}$ and $C$ is the marginalised parameter covariance, and the subscripts BOSS and ${\it Planck\/}$ refer to quantities estimated by our BOSS analysis and by {\it Planck\/}, respectively. In this three dimensional space, we find a 2.3$\sigma$ tension between {\it Planck\/} and BOSS, which remains at the same level in our ``Physical Bias Prior'' analysis (a decrease in the tension in $h$ owing to SGC data is compensated by increased tensions in $\Omega_{\rm m}$ and $\sigma_8$).

To explore this tension further, we reanalysed the BOSS data but this time fixing the cosmological parameters to the values preferred by {\it Planck\/}. This will tell us whether there exists a set of bias and nuisance parameters that provide a good fit to BOSS within the {\it Planck\/} cosmology.

Indeed, in Fig.~\ref{fig:PlanckFix} we see that {\it Planck\/} and our best-fit cosmology both provide a good description of the low and high-$z$ BOSS data, especially at $k < 0.25\,\ihMpc$. Although not shown here, the two parameters showing the greatest deviations when fixing to this new cosmology are $b_1$ and $b_{s^2}$ (which are also more tightly constrained). Specifically, the linear bias $b_1$ decreases to accommodate for a higher value of $\sigma_8$, while the tidal field bias parameter shows a preference towards positive non-zero values, $b_{s^2}\simeq1.3$ at the 2-$\sigma$ confidence level. Note that from our studies in \cite{Zennaro_2022}, it is not possible to completely rule out the possibility for BOSS galaxies to live in regions where the Lagrangian tidal field is positive, specially given the complex selection criteria of BOSS and that these bias relations were built for a generic galaxy formation sample.

The value of $\chi^2$ increases only mildly when the cosmology is fixed to {\it Planck\/}, from $446.03$ to $448.02$. From a frequentist point of view, the difference of $\Delta \chi^2=2$ is expected to occur 60\% of the times considering that we have $456$ data points, minus some allowance for the nonlinear degrees of freedom in the model, which we neglect. Thus apparently the {\it Planck\/} cosmology is in good agreement with the BOSS data when looked at in this way. This may however partly reflect the fact that the $\chi^2$ test is something of a blunt tool: it lacks knowledge of the form of the model under study, which is an explicit feature of the Bayesian analysis.


Regarding the degree to which the model's non-cosmological parameters need to be fine-tuned to accommodate {\it Planck\/}, we can assess this by computing the Bayesian evidence ratio, $R$, defined as the ratio between the evidences of a model adopting the {\it Planck\/} set of values and our BOSS cosmology. The result is $\ln R \simeq -0.5$, which is well bellow the commonly accepted threshold of $4.6$ for a decisive preference for one model over another \citep{jeffreys1998theory}. Hence, we conclude that the full-shape BOSS clustering is in no significant tension with the {\it Planck\/} set of cosmological parameters from the Bayesian point of view, despite the decrease in number of dimensions.

Finally, as shown in dotted lines in Fig.~\ref{fig:PlanckFix}, we can repeat this comparison using our ``Physical Bias Prior" analysis. By using the {\it Planck\/} cosmological parameters and further fixing the high-order bias parameters to their ``Physical Bias Prior" values, we find a 1-$\sigma$ distance from the fiducial result in the best-fit models, $\Delta \chi^2 = 15$, and a Bayes ratio of $\ln R \simeq -8.2$, indicating a decisive evidence in favour of this case over that where cosmology is varied, despite the larger $\chi^2$. This result can be interpreted as follows: the nuisance parameters of our model can accommodate a fit where the cosmological parameters are set to {\it Planck\/} and where the ``Physical Bias Prior" are used. This fit is only partially excluded, since it is only 1-$\sigma$ away from the fiducial result. However, since the dimensionality reduced from 21 to 10, the Bayes factor heavily penalises the inclusion of these extra free parameters. 

A fairer test is to compare cases where the cosmology is either fixed to {\it Planck\/} or to the best-fit BOSS-NGC cosmology. This gives a Bayes ratio of $\ln R \simeq 2.1$, providing substantial evidence for the BOSS cosmology over Planck, when assuming ``Physical Bias Priors". We caution, however, that we do not expect BOSS galaxies to exactly coincide with our bias priors. Instead, this result highlights the benefits of adopting informative bias priors: although our cosmological constraints did not improve significantly after adopting relationships among bias parameters, such priors can be extremely valuable when comparing competing models. Therefore, we emphasise the future need for robust priors on nuisance parameters, estimated according to our state-of-the-art understanding of galaxy formation physics.





\section{Conclusions}
\label{sec:conclusions}

In this paper, we have presented a novel analysis of the multipoles of the redshift-space power spectrum of galaxies in the SDSS-IV BOSS survey. We employed a novel theoretical model based on a perturbative description of the relationship of matter and galaxies, which we combined with displacement and velocity fields extracted from cosmological simulations. The advantages of our novel modelling allow us to employ much smaller scales ($\kmax \simeq 0.4\ihMpc$) than those typically used previous analyses. 

Our fiducial analysis of the NGC provides one of the most accurate measurements to date on $\Omega_{\rm m}$, $h$, and $\sigma_{8}$ from the BOSS dataset multipoles alone. Specifically, we obtain an accuracy of 3.6\% on $\Omega_m=0.301\pm 0.011$, 4\% on $\sigma_8=0.745^{+0.028}_{-0.035}$, 2\% on  $h=0.705\pm 0.015$, and 4\% accuracy on the structure parameter, $S_8 = 0.747^{+0.032}_{-0.039}$. 

Our constraint on $S_8$ is $2.5\sigma$ lower than the value inferred by the analysis of the {\it Planck\/} satellite data. This low $S_8$ value is compatible with constraints from other low-$z$ probes such as recent analyses of the combination of gravitational shear, angular clustering, and their cross correlation. However, we argue that the current BOSS data are not sufficient to confidently identify a tension with {\it Planck\/}. Specifically, we showed that the {\it Planck\/} cosmology is still able to achieve a good fit to the power spectrum data, and the differences with respect to our best-fit cosmology can be absorbed by the astrophysical nuisance parameters of our model.

Another novel aspect of our analysis was the adoption of physical priors on the relationships between the linear and higher order bias parameters. Adopting these priors reduces prior volume effects and increases the accuracy of our fiducial cosmological constraint on $S_8$ by $10\%$. However, we note that our priors correspond to the mean expected in a generic galaxy sample, and were not built specifically for a BOSS-like sample. In this sense, this analysis can be seen as exploring the potential gains and benefits of combining flexible models with our information arising from our current understanding of galaxy formation. 

In the future, by considering only specific astrophysical scenarios appropriate to the particular galaxy sample under study, we expect to obtain more robust and even more informative priors on the nuisance parameters of our model. Importantly, this approach provides a clear and systematic route to incorporate state-of-the-art galaxy formation simulations into cosmological inference. We plan to explore this in forthcoming publications.

 
\section*{Acknowledgements}

MPI is supported by STFC consolidated grant no. RA5496. REA acknowledges support from project PID2021-128338NB-I00 from the Spanish Ministry of Science and support from the European Research Executive Agency HORIZON-MSCA-2021-SE-01 Research and Innovation programme under the Marie Skłodowska-Curie grant agreement number 101086388 (LACEGAL). We would like to thank Oliver Philcox for providing with the Patchy-Nseries data, Naomi Robertson, Martin White, Giovanni Arico, David Alonso, Matteo Zennaro, Alex Amon, Masahiro Takada, and Pedro Gregorio Carrilho for useful discussions. For the purpose of open access, the author has applied a Creative Commons Attribution (CC BY) licence to any Author Accepted Manuscript version arising from this submission.

\section*{Data Availability}

This work used data of the BOSS-SDSS collaboration\footnote{Available at \url{https://data.sdss.org/sas/dr12/boss/lss/}}, specifically, in its window-free power spectra version\footnote{Available at \url{https://github.com/oliverphilcox/Spectra-Without-Windows}}. It also made extensive use of the \nseries mock catalogues\footnote{Available at \url{https://www.ub.edu/bispectrum/page11.html}}. The main theory package used in this work is the \texttt{baccoemu} package\footnote{Available at \url{https://baccoemu.readthedocs.io/en/latest/}}, and for the MCMC chains, the \texttt{MULTINEST}\footnote{Available at \url{https://github.com/farhanferoz/MultiNest}} code.



\bibliographystyle{mnras}
\bibliography{example} 




\appendix

\section{Appendix A}

In this appendix, we present the full marginalised posteriors for all cosmological and non-cosmological parameters sampled within the MCMC likelihoods, using the power spectrum multipoles. We display them for the three cases shown in Fig.~\ref{fig:MoneyPlot}, namely using the BOSS-NGC, BOSS-SGC, and combined BOSS datasets with physical bias priors.

\begin{figure*}
\includegraphics[width=\textwidth]{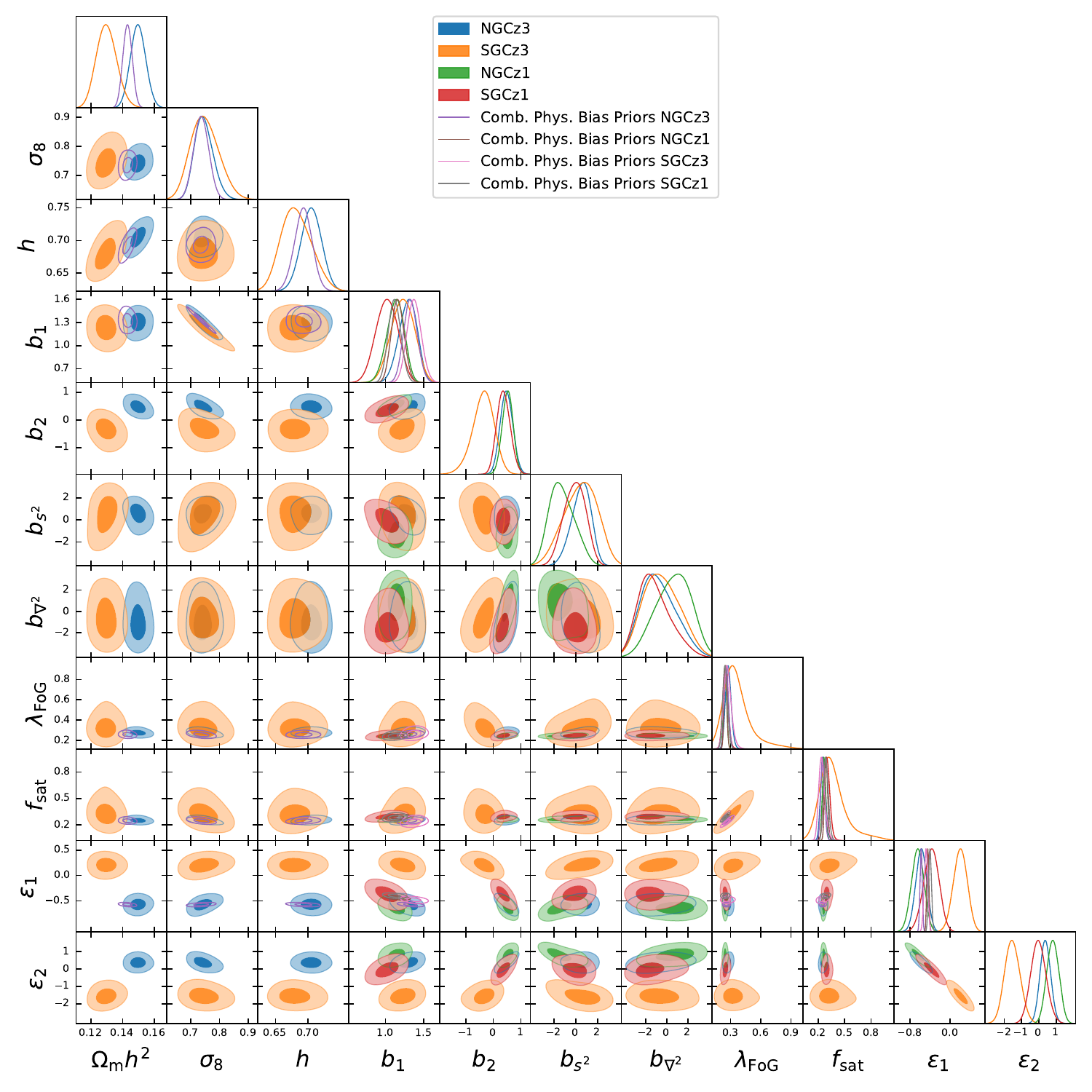}
\caption{Marginalised constraints on the full set of parameters of our model. We show results for the separate fiducial analysis of NGC and SGC, and for their combination adopting physical bias priors. In the case of nuisance parameters, we display the respective constraints for each subsample.}
\label{fig:FullPosterior}%
\end{figure*}


\bsp	
\label{lastpage}
\end{document}